\documentclass[twoside]{article}
\usepackage{qic,epsfig,epstopdf,float,placeins, textgreek}

\begin{document}
\setlength{\textheight}{8.0truein}    

\runninghead{Fast quantum modular exponentiation architecture for Shor's factoring alogrithm}
{A. Pavlidis and D. Gizopoulos}

\normalsize\textlineskip
\thispagestyle{empty}
\setcounter{page}{649}

\copyrightheading{14}{7\&8}{2014}{0649--0682}

\vspace*{0.88truein}

\alphfootnote

\fpage{649}

\centerline{\bf
FAST QUANTUM MODULAR EXPONENTIATION ARCHITECTURE}
\vspace*{0.035truein}
\centerline{\bf FOR SHOR'S FACTORING ALGORITHM}
\vspace*{0.37truein}
\centerline{\footnotesize
ARCHIMEDES PAVLIDIS}
\vspace*{0.015truein}
\centerline{\footnotesize\it Department of Informatics and Telecommunications, National and Kapodistrian University of Athens}
\baselineskip=10pt
\centerline{\footnotesize\it Panepistimiopolis, Ilissia, GR 157 84, Athens, Greece}
\baselineskip=10pt
\centerline{\footnotesize\it University of Piraeus, 80, Karaoli \& Dimitriou St., GR 185 34, Piraeus, Greece}
\baselineskip=10pt
\centerline{\footnotesize\it e-mail: adp@unipi.gr}
\vspace*{10pt}
\centerline{\footnotesize 
DIMITRIS GIZOPOULOS}
\vspace*{0.015truein}
\centerline{\footnotesize\it Department of Informatics and Telecommunications, National and Kapodistrian University of Athens}
\baselineskip=10pt
\centerline{\footnotesize\it Panepistimiopolis, Ilissia, GR 157 84, Athens, Greece}
\baselineskip=10pt
\centerline{\footnotesize\it e-mail: dgizop@di.uoa.gr}
\vspace*{0.225truein}
\publisher{November 28, 2012}{October 2, 2013}

\vspace*{0.21truein}

\abstracts{
We present a novel and efficient, in terms of circuit depth, design for Shor's quantum factorization algorithm. The circuit effectively utilizes a diverse set of adders based on the Quantum Fourier transform (QFT) Draper's adders to build more complex arithmetic blocks: quantum multiplier/accumulators by constants and quantum dividers by constants. These arithmetic blocks are effectively architected into a quantum modular multiplier which is the fundamental block for the modular exponentiation circuit, the most computational intensive part of Shor's algorithm. The proposed modular exponentiation circuit has a depth of about $2000n^2$ and requires $9n+2$ qubits, where $n$ is the number of bits of the classic number to be factored. The total quantum cost of the proposed design is $1600n^3$. The circuit depth can be further decreased by more than three times if the approximate QFT implementation of each adder unit is exploited. 
}{}{}

\vspace*{10pt}

\vspace*{3pt}
\communicate{R~Cleve~\&~B~Terhal}

\vspace*{1pt}\textlineskip    
\section{Introduction}\label{sec:Introduction}     
\noindent   
One of the most well-known quantum algorithms is Shor's algorithm \cite{Shor} for integer factorization. It is currently one the most promising algorithm for implementation on a quantum computer due to its extremely important applicability in the cryptanalysis field. Compared to its classical (non-quantum) number factorization counterpart algorithms, Shor's factorization algorithm offers superpolynomial execution speedup.

The quantum part of Shor's algorithm essentially consists of the two main blocks shown in Figure \ref{fig:HighLevelShor}: 

\begin{figure} [ht]
\centerline{\epsfig{file=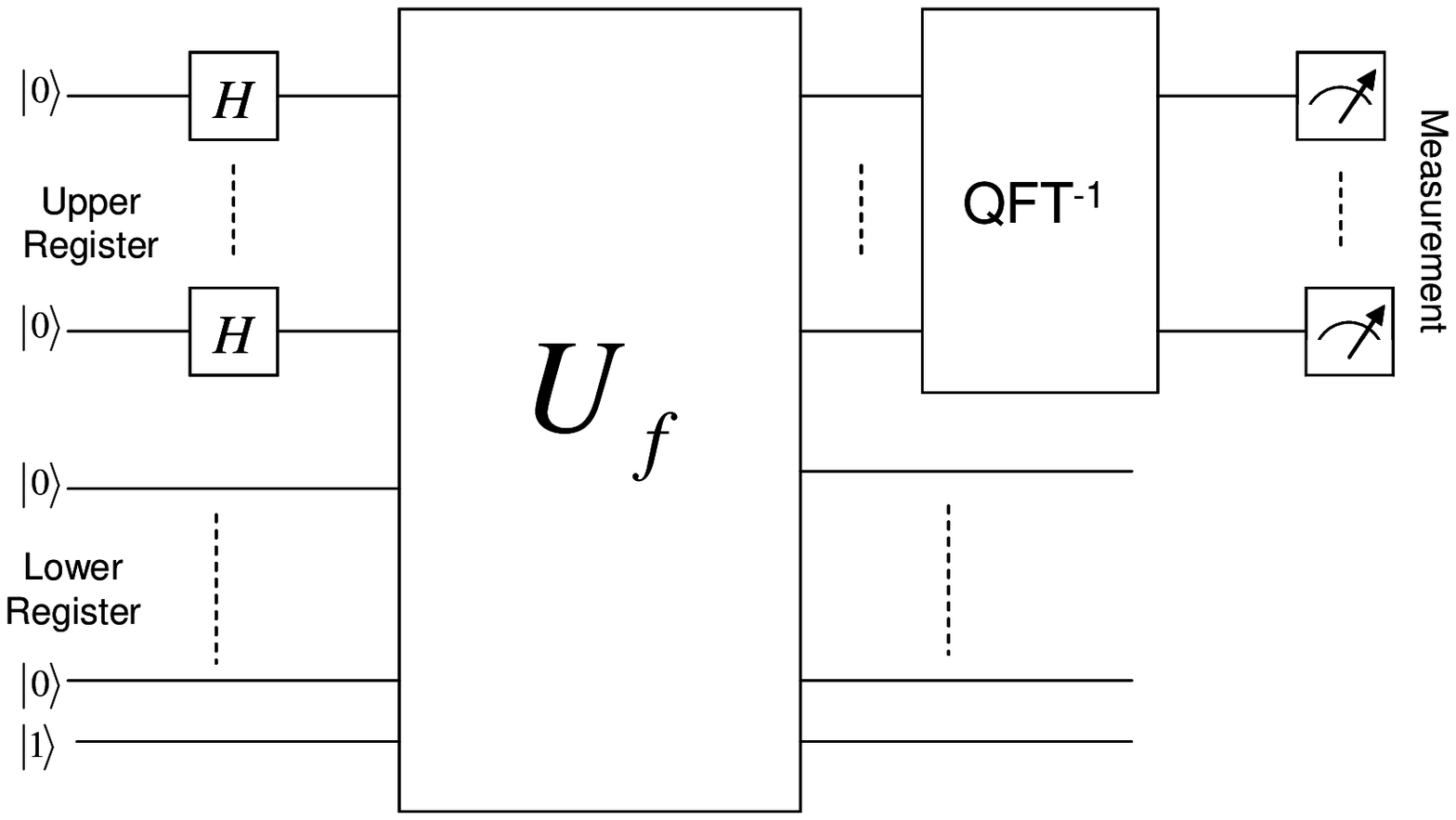, width=8.2cm}} 
\vspace*{13pt}
\fcaption{\label{fig:HighLevelShor}High level diagram of Shor's algorithm. Upper register consists of $2n$ qubits and holds the superposition of integers $0 \ldots N^2-1$; lower register consists of $n$ qubits and holds the superposition of values $a^{x} \bmod N$ after computed by $U_{f}$ block. Classical postprocessing of the measurement in the computational basis after the QFT block  gives with high probability the period of the function $f(x) = a^{x} \bmod N$.}
\end{figure}

\begin{itemize}
\item 
A modular exponentiation circuit $U_{f}$ computing the function $f(x) = a^x \bmod N$, where $N$ is the $n$-bit number to be factored, $a$ is a random number between 2 and $N-1$ which is co-prime to $N$, and $x$ is the argument of the function taking values between $0$ and at least $N^2$.
\item
An inverse Quantum Fourier Transform (QFT\textsuperscript{-1}) circuit. 
\end{itemize}

The most computational intensive quantum part of Shor's algorithm is the modular exponentiation circuit $U_{f}$. It applies the following transform to its quantum input:
\begin{equation}
U_{f}=(|x\rangle |1\rangle)=|x\rangle |a^x \bmod N\rangle.
\label{eq:modexp}
\end{equation}

The purpose of the Hadamard ($H$) gates at the upper register is to create a quantum superposition of numbers $x$ ($x=0 \ldots N^2$) before providing it to the modular exponentiation circuit $U_{f}$. The measurement gates jointly give an integer which may give the factors of $N$ with high probability through further classical post-processing. If not succeeding then Shor's algorithm is re-executed with a new random number $a$.

Our work proposes an architecture for the modular exponentiation section of the Shor's  algorithm which consists of the following main components:

\begin{itemize}
\item
a quantum controlled multiplier/accumulator by constant, based on previous work done by Draper \cite{Draper1} and Bauregard \cite{Beauregard}, but having the advantage of linear depth exploiting the fact that one of the factors is constant, decomposing doubly controlled rotation gates to simple controlled rotation gates \cite{Barenco1} and then suitably rearranging them, and 
\item
a quantum divider by constant based on the division by constant algorithm proposed by Granlund and Montgomery \cite{Granlund}. This quantum divider also has a linear depth. By combining the above building blocks we achieve to build a quantum controlled modular multiplier with linear depth which is then used as the basic building block for the quantum modular exponentiation circuit of quadratic depth. This modular exponentiation circuit has same depth order and qubits order with circuits proposed by Zalka \cite{Zalka} and Kutin \cite{Kutin}, but it computes the exact result, while the last two mentioned circuits operate in a way that results to approximate computations.
\end{itemize}

This paper is organized as follows. Section \ref{sec:Background} provides a background for the design of modular exponentiation circuits from elementary quantum arithmetic circuits, summarizes related work and discusses previously published arithmetic blocks that will be used in the proposed design. Section \ref{sec:FMAC} presents a linear depth multiplier by constant and accumulate unit based on QFT. Section \ref{sec:GMFDIV} introduces a linear depth divider by constant unit, also based on QFT. In Section \ref{sec:FMOD1}, we combine the two previous units and describe a generic version of a modular multiplier which can be used as the building block of a modular exponentiation circuit. In Section \ref{sec:FMOD2} an optimized version of a modular multiplier is presented. In Section \ref{sec:Analysis} we give a detailed complexity analysis both in terms of space and time along with comparisons with other circuits presented in the literature; implementation difficulties are also discussed. Section \ref{sec:Conclusion} concludes the paper. 

\section{Background}\label{sec:Background}
\noindent

\subsection{Decomposition into modular multipliers}
\noindent
If $\lceil log_{2}N \rceil$  is the required number of bits to represent the number $N$ to be factored, we can see that the upper quantum register in Figure \ref{fig:HighLevelShor} requires at least $2n$ qubits, because Shor's algorithm requires $x$ to take values between 0 and at least $N^2$. It also requires $n$ qubits for the lower quantum register (to hold a modulo $N$ integer), leading to a total requirement of $3n$ qubits. The modular exponentiation function can be written as:

\begin{equation}
a^{x} \bmod N = (a^{2^{0}} \bmod N)^{x_{0}} \cdot (a^{2^{1}} \bmod N)^{x_{1}} \cdots (a^{2^{2n-1}} \bmod N)^{x_{2n-1}}  \bmod N
\label{eq:modexpfactors}
\end{equation}

where $x_{i},  (i=0 \ldots 2n-1)$ are the bits of the binary expansion of $x$, that is   $x=(x_{2n-1},\ldots,x_{1},x_{0})$   and we can readily produce the equivalent design shown in Figure \ref{fig:HighLevelShorDecomposed}.

This design uses $2n $ controlled-$U$ ( $CU$) blocks. Each block denoted as $CU_{a^{2^{i}}}$ is a controlled modular $N$ multiplier of its input (the lower quantum register) by constant $a^{2^{i}} \bmod N$. Each block is controlled by the corresponding 
$x_{i}$ qubit of the upper quantum register. Hence, it generates a transformation on    $n + 1$ qubits of the form ($c$ is the control qubit and $y$ is the input qubit):

\FloatBarrier

\begin{figure} [!hb]
\centerline{\epsfig{file=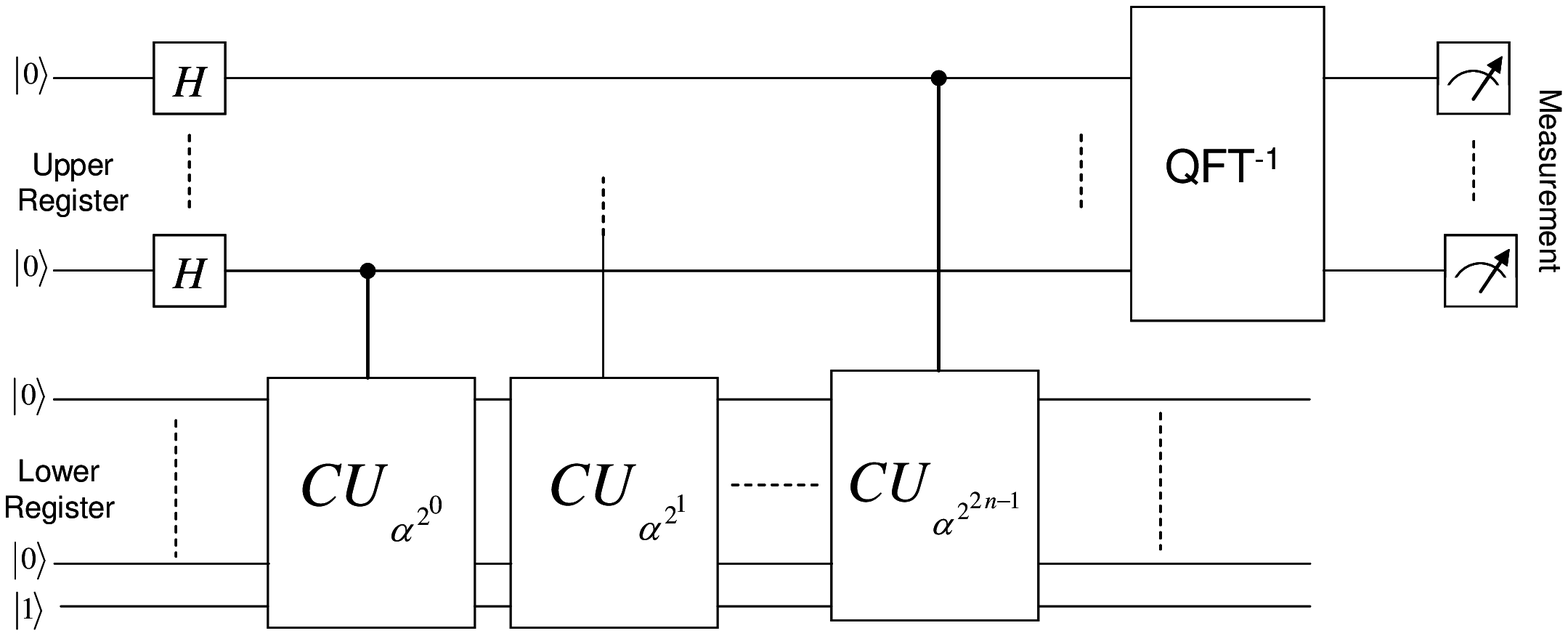, width=12cm}} 
\vspace*{13pt}
\fcaption{\label{fig:HighLevelShorDecomposed} Design of modular exponentiation circuit using controlled modular multipliers. Each multiplier performs a modular multiplication   $ | ay \bmod N \rangle $ on the lower register when its controlling qubit is equal to    $ | 1 \rangle $  .}
\end{figure}

\begin{figure} [!ht]
\centerline{\epsfig{file=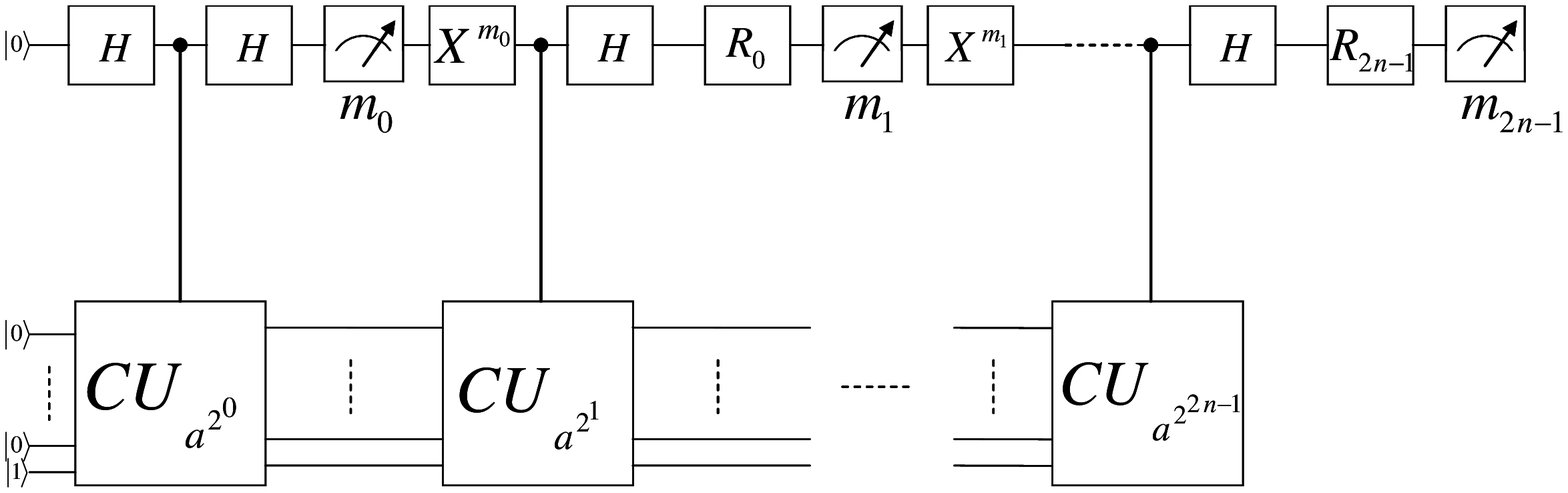, width=12cm}} 
\vspace*{13pt}
\fcaption{\label{fig:HighLevelShor1qubit} Design of modular exponentiation circuit using only one qubit to control the modular multipliers. The phase shift gates $R$ depend on all previous measurement results and implement the inverse QFT, while the $X$ gates are negations conditioned on the result of each measurement \cite{Beauregard}.}
\end{figure}

\FloatBarrier

\begin{equation}
CU_{a^{2^{i}}}(|c\rangle |y\rangle)=|c\rangle |(a^{2^{i}})^{c}y \bmod N\rangle
\label{eq:CUblock}
\end{equation}

The $CU_{a^{2^{i}}}$ blocks commute and the inverse QFT can be implemented semiclassically. Therefore, the circuit can be re-designed into that of Figure \ref{fig:HighLevelShor1qubit} which uses only one qubit for controlling $2n$ $CU_{a^{2^{i}}}$ gates instead of $2n$ different qubits (\cite{Beauregard}, \cite{Mosca}, \cite{Parker}).

\subsection{Previous work on Shor's algorithm designs}
\noindent

It is well-known that the most computational intensive part of a quantum circuit for Shor's algorithm is the modular exponentiation circuit. Several designs for quantum exponentiation have been proposed in the literature. In general, they adopt a top-down approach where the modular exponentiation is realized using \emph{modular multiplier} blocks, which in turn are constructed using \emph{modular adder} blocks, then using the lowest level block, the \emph{adder}. For this reason, most of the effort has been devoted to the design of the quantum equivalent of a digital adder and its improvement in terms of complexity: reduction of the total number of required qubits (ancilla and working) and reduction of the depth of the circuit (total number of steps required to complete the computation). 
Many of the quantum addition circuits introduced in the literature are inspired from their known classical counterparts through the design of reversible versions of them. The most important of these approaches are the Vedral-Barenco-Ekert (VBE) ripple-carry adder \cite{Vedral}, the Beckman-Chari-Devabhaktuni-Preskill (BCDP) ripple-carry adder \cite{Beckman}, the Cuccaro-Draper-Kutin-Moulton (CDKM) ripple carry adder \cite{Cuccaro}, the Draper-Kutin-Rains-Svore (DKRS) carry-lookahead adder \cite{Draper2}, the Takahashi-Kunihiro (TK) ripple-carry adder \cite{Takahashi}, the Gossett carry-save adder \cite{Gossett}, the Zalka carry-select adder \cite{Zalka} and the VanMeter-Itoh (VI) carry-select adder and conditional-sum adder \cite{VanMeter1}. Some of the above proposals for quantum addition circuits emphasize on minimizing on the number of required qubits, while other methods try to minimize the depth of the circuit. Other efforts concentrate on building architectures restricted on the condition of local communications between the qubits either in 1D-NTC (1-Dimension, linear Nearest-neighbour, Two qubit gates, Concurrent execution) such as those of Fowler-Devitt-Hollenberg (FDH) \cite{Fowler1} and Kutin's \cite{Kutin}, or in 2D-NTC  such as those of Choi-VanMeter (CV) \cite{Choi1} and Pham-Svore (PS) \cite{Pham}.
The method to build a complete exponentiation circuit based on a particular addition circuit is not unique, and various studies concerning the trade-off between space (number of required qubits) and time (depth of the circuit) have been reported as in \cite{VanMeter1}. Not all previous publications provide a complete modular exponentiation circuit, but assuming we can build one using the previously discussed hierarchy (adder)-(modular adder)-(modular multiplier)-(modular exponentiator), we can make rough approximations about the design complexity in each case and comparisons with the proposed exponentiation design (see Section \ref{sec:Analysis} for the comparisons). 
Notable exceptions of the above top-down trend that builds a complete modular exponentiation circuit from the quantum equivalent of classical binary adder are circuits that use the Draper's QFT adder \cite{Draper1} like Beauregard's circuit \cite{Beauregard}, Fowler-Devitt-Hollenberg circuit  (FDH)  \cite{Fowler1}, and Kutin's first circuit of \cite{Kutin} . These three circuits implement the addition of two integers by converting one of them in the Fourier domain using QFT and then converting the sum back to the binary representation. Another method which surpassess the common hierarchy of computation is Zalka's FFT multiplier \cite{Zalka}  that implements a multiplier using the FFT method of computing a convolution sum. 
Our novel design of the quantum modular exponentiation architecture concentrates mainly on minimizing the circuit depth. It adopts the Draper's QFT adder \cite{Draper1}  as a basic building block, modifies Beauregard's multiplier/accumulator \cite{Beauregard}  so as to reduce its depth from $O(n^{2})$ to $O(n)$, and by developing a quantum divider based on Granlund-Montgomery classical division by constant algorithm \cite{Granlund}, it hierarchically builds upon the following sequence: (adder) – (multiply/accumulator) – (constant divider) – (modular multiplier) – (modular exponentiator).

\subsection{QFT adders – {\textbf{\textit \textPhi}}ADD, C{\textbf{\textit \textPhi}}ADD, CC{\textbf{\textit \textPhi}}ADD}
\begin{figure} [!ht]
\centerline{\epsfig{file=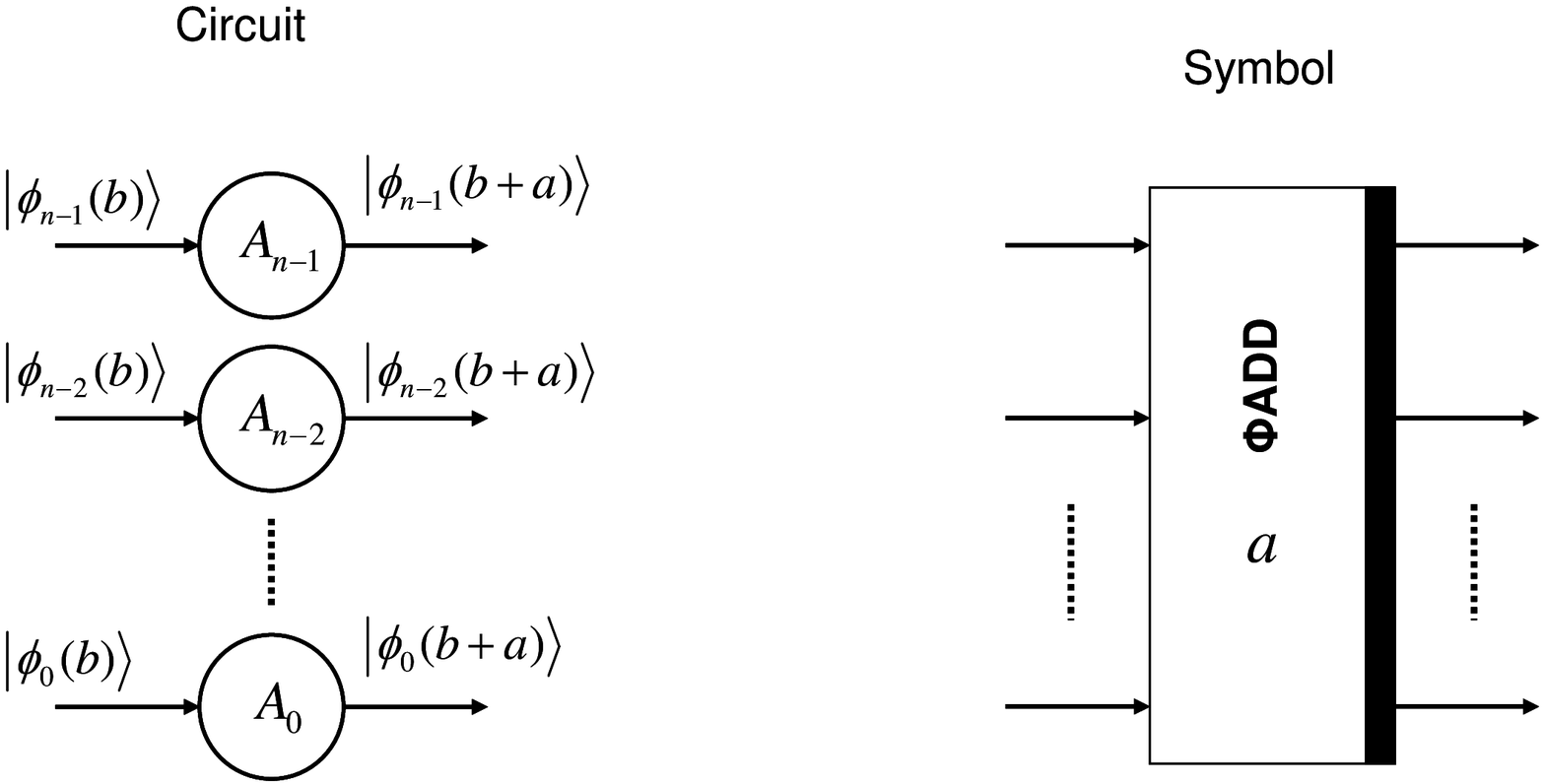, width=9cm}} 
\vspace*{13pt}
\fcaption{\label{fig:FADD} $\Phi$ADD adder circuit of depth 1 and its symbol. This circuit adds a constant integer $a$ to the quantum integer $b$, when $b$ is already in the Fourier domain. The value of integer $a$ is hardwired in the angles of the phase shift gates $A_{j}, j=0 \ldots n-1$ as defined in Eq. \ref{eq:AdderAngles}.}
\end{figure}

\noindent
We first describe the four QFT adders \cite{Draper1} that will be extensively used in our modular exponentiation design. Since in every iteration of Shor's algorithm the randomly picked number $a$ in Eq. \ref{eq:modexp} remains constant, we need an adder receiving a quantum integer (that is a potential superposition of integer $x$ as required by the algorithm) as its first operand and a constant classical integer as its second. Three variations of this adder are required for the multiplier unit: the QFT constant adder ($\Phi$ADD), the controlled QFT constant adder (C$\Phi$ADD) and the doubly controlled QFT constant adder (CC$\Phi$ADD). A generic QFT adder for adding two quantum integers will be also used subsequently in the quantum divider circuit. These four QFT adders will be denoted in all figures as $\Phi$ADD and they will be easily differentiated by the quantum wires connected as inputs and outputs on their symbols.
Figure \ref{fig:FADD} shows the simple (uncontrolled) QFT constant adder $\Phi$ADD and its symbol. It adds the $n$-bit constant integer $a$ to an $n$-qubit quantum number        $|b \rangle = |b_{n-1} \rangle  \cdots  |b_{1} \rangle |b_{0} \rangle = |b_{n-1} \rangle \otimes \cdots \otimes |b_{1} \rangle |b_{0} \rangle $. Number $b$ must be already transformed in the Fourier domain by a QFT block before entering the $\Phi$ADD block through the relation:

\begin{equation}
|b \rangle \stackrel{{\scriptscriptstyle QFT}}{\longrightarrow} |\phi (b) \rangle = |\phi_{n-1} (b) \rangle |\phi_{n-2} (b) \rangle \ldots |\phi_{0} (b) \rangle =   \frac{1}{\sqrt{2^{n}}} \sum_{k=0}^{2^{n}-1}e^{i \frac{2 \pi}{2^{n}} bk} |k \rangle
\label{eq:QFT}
\end{equation}

The individual $j^{th}$ qubit $|\phi_{j} (b) \rangle$ of the quantum Fourier transformed number is given by the relation:

\begin{equation}
|\phi_{j} (b) \rangle =  \frac{1}{\sqrt{2}}  ( |0 \rangle + e^{i \frac{2 \pi}{2^{j}} b}      |1 \rangle  )
\label{eq:QFTqubit}
\end{equation}

\FloatBarrier

The one-qubit gates shown in Figure \ref{fig:FADD} are rotation quantum gates each one described by the equations ($R_{k}$ is a phase shift gate):

\begin{equation}
\begin{array}{lr}
 A_{j}= \prod\limits_{k=1}^{j+1} R_{k}^{a_{j+1-k}},
&

R_{k}=   \left[
\begin{array}{cc}
1 & 0 \\
0 & e^{i \frac{2 \pi}{2^{k}}}
\end{array} \right]

\end{array}
\label{eq:AdderAngles}
\end{equation}

Therefore, it can be shown that the output of the $\Phi$ADD circuit is $|\phi (b+a) \rangle $, the quantum Fourier transformed sum $b+a$. The sum can be recovered in the computational basis after applying an inverse QFT. Excluding the forward and inverse QFT, the circuit has a complexity of $n$ qubits and a depth of 1 because the rotation gates can all operate in parallel.

\begin{figure} [!hb]
\centerline{\epsfig{file=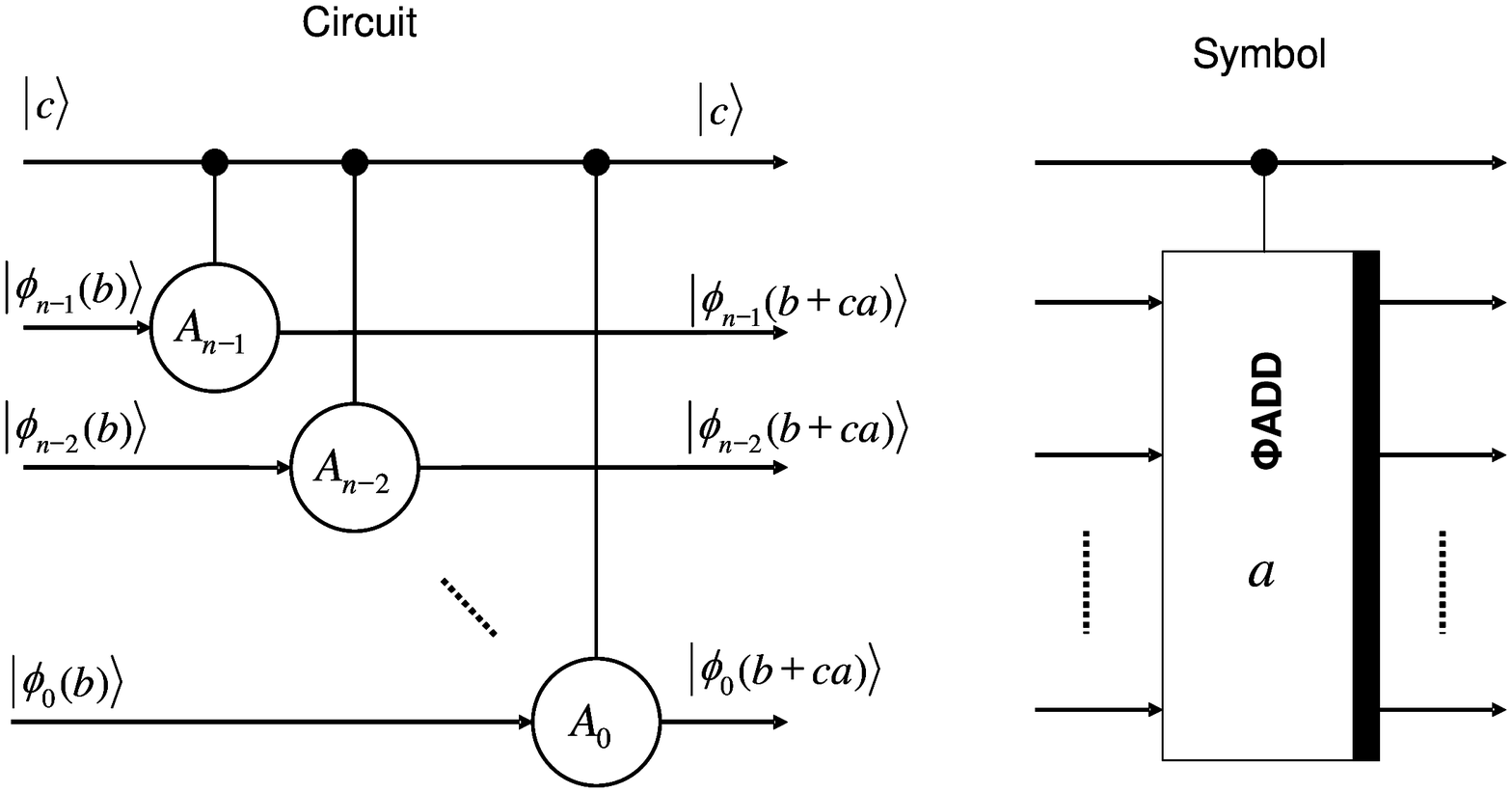, width=9cm}} 
\vspace*{13pt}
\fcaption{\label{fig:CFADD} C$\Phi$ADD controlled adder circuit of depth $n$ and its symbol. This circuit adds the constant $a$ to the quantum integer $b$ when the control qubit $|c \rangle$ is $|1 \rangle$. Again, the constant value $a$ is hardwired in the controlled rotation gates as defined in Eq. \ref{eq:AdderAngles}.  }
\end{figure}

An extension of the constant adder $\Phi$ADD circuit can be done if we use controlled rotation gates with same rotation angles as those defined in Eq. \ref{eq:AdderAngles}. This circuit and its symbol are depicted in Figure \ref{fig:CFADD}. It has a common controlling qubit $|c \rangle $ for each rotation gate and performs the following transform

\FloatBarrier

\begin{figure} [!ht]
\centerline{\epsfig{file=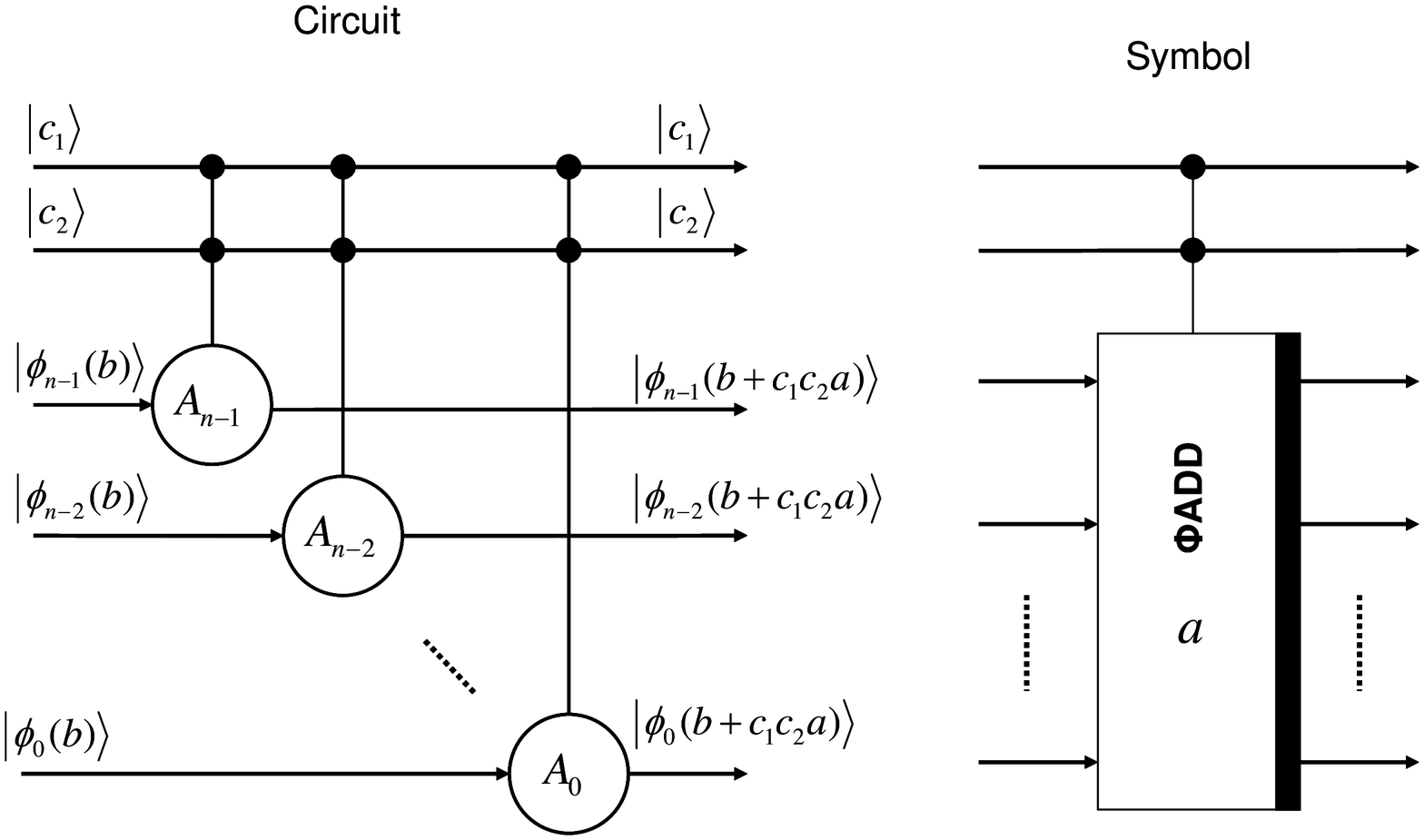, width=9cm}} 
\vspace*{13pt}
\fcaption{\label{fig:CCFADD} The doubly-controlled adder circuit CC$\Phi$ADD of depth $n$ and its symbol. This is an extension of the C$\Phi$ADD circuit and the addition is performed when both the control qubits $|c_{1} \rangle$ and $|c_{2} \rangle$ are $|1 \rangle$.  }
\end{figure}

\begin{equation}
C\mathit{\Phi}ADD _{a}( |c \rangle |\phi(b) \rangle)= |c \rangle |\phi(b+ca) \rangle
\label{eq:CFADD}
\end{equation}

The C$\Phi$ADD circuit performs the addition only when the controlling qubit $|c \rangle$ is $|1 \rangle$  giving the result $|\phi(b+ca) \rangle$, otherwise the result is the input  $|\phi(b) \rangle$.The C$\Phi$ADD adder needs $n + 1$ qubits and has a depth of $n$ because the controlled rotation gates must operate sequentially as they have a common controlling qubit.

A further extension is shown in Figure \ref{fig:CCFADD}, which is the doubly-controlled $\Phi$ADD circuit (CC$\Phi$ADD) and its symbol. The circuit is similar to the C$\Phi$ADD, but its gates are doubly-controlled rotation gates. The two controlling qubits of each rotation gate are $|c_{1} \rangle$ and $|c_{2} \rangle$. The CC$\Phi$ADD circuit performs the transform:

\FloatBarrier

\begin{figure} [H]
\centerline{\epsfig{file=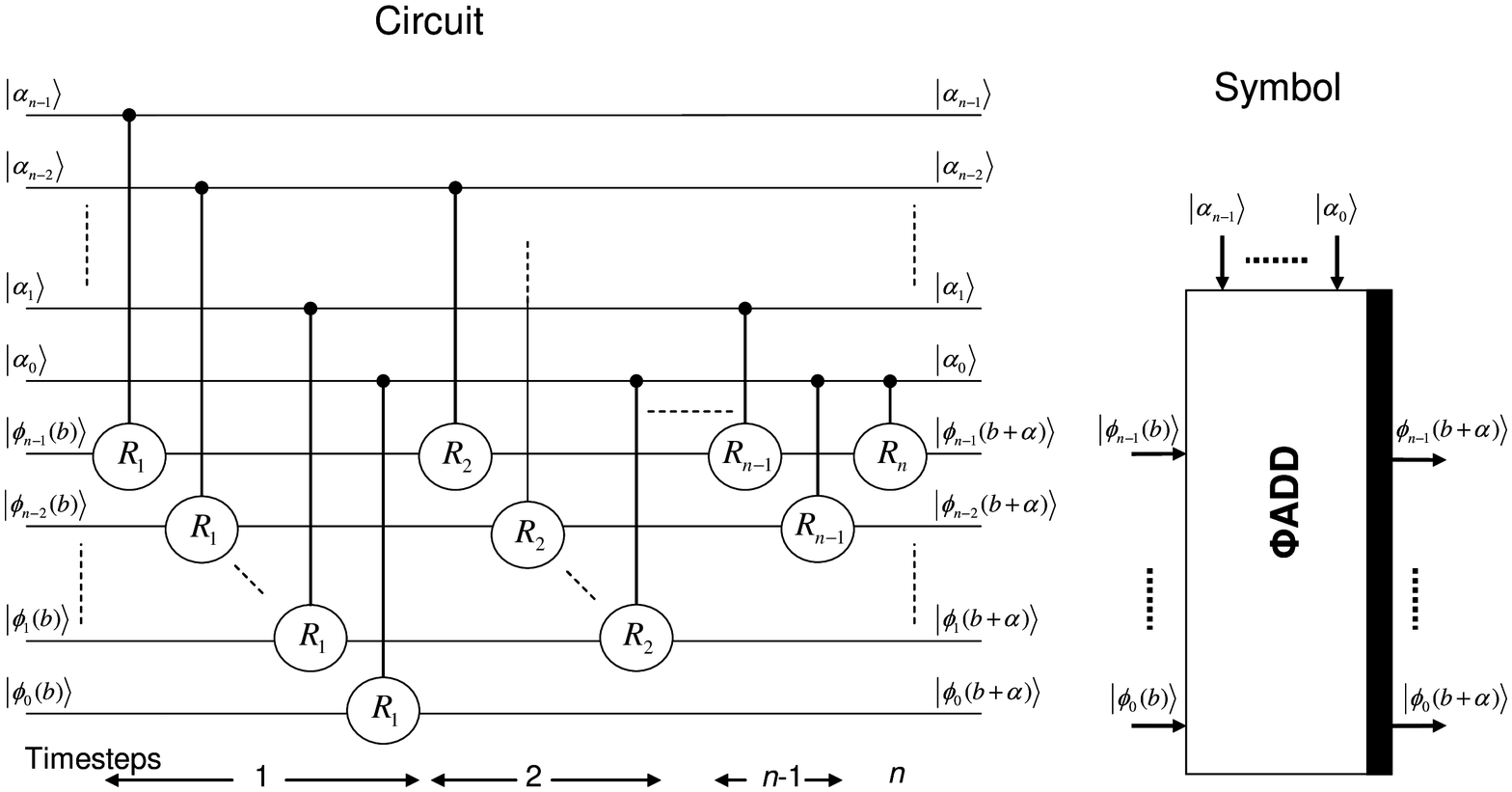, width=12cm}} 
\vspace*{13pt}
\fcaption{\label{fig:GFADD} Generic adder $\Phi$ADD circuit and its symbol. The upper input bus in the symbol consists of the qubits $|a_{0} \rangle, \ldots ,|a_{n-1} \rangle$ that control the rotation gates.  }
\end{figure}

\FloatBarrier

\begin{equation}
CC\mathit{\Phi}ADD _{a}( |c_{1} \rangle |c_{2} \rangle |\phi(b) \rangle)= |c_{1} \rangle |c_{2} \rangle |\phi(b+c_{1}c_{2}a) \rangle
\label{eq:CCFADD}
\end{equation}

That is, it performs the addition only when both the controlling qubits $|c_{1} \rangle$ and $|c_{2} \rangle$ are $|1 \rangle$ and results  $|\phi(b+a) \rangle$, otherwise the result is the input $|\phi(b) \rangle$. The CC$\Phi$ADD adder needs $n + 2$ qubits and like the C$\Phi$ADD has a depth of $n$ because the doubly-controlled rotation gates must operate sequentially as they have common controlling qubits.

Finally, we give in Figure \ref{fig:GFADD} the circuit diagram and symbol of the generic QFT adder ($\Phi$ADD) that adds two quantum numbers. The circuit diagram of Figure 7 is the parallel version of the adder that has a depth of $n$. The operation of the circuit is to add two quantum integers each of $n$ qubits and is described by Eq. \ref{eq:GFADD}: 

\begin{equation}
\mathit{\Phi}ADD( |a \rangle |\phi(b) \rangle)= |a \rangle |\phi(b+a) \rangle
\label{eq:GFADD}
\end{equation}

The upper input bus in the symbol consists of the qubits $|a_{0} \rangle, \ldots ,|a_{n-1} \rangle$ that control the rotation gates.. These qubits remain unaltered by the $\Phi$ADD unit.
As shown in Figure \ref{fig:GFADD} and Eq. \ref{eq:GFADD} one of the integers must be already transformed in the QFT domain which will also be the case for the sum result.

\subsection{Fourier Multiplier/Accumulator - {\textbf{\textit \textPhi}}MAC}
\noindent
In this section we describe a quantum circuit given in \cite{Beauregard}, which from this point onwards we name $\Phi$MAC; it utilizes the CC$\Phi$ADD adders described in the previous subsection and accumulates the product of a constant $n$-bit integer $a$ with a quantum $n$-qubit integer, $|x \rangle$ to a quantum $2n$-qubit integer $|b \rangle$, giving the accumulation result $|b+ax \rangle$. Furthermore, the circuit has a controlling qubit $|c \rangle$, that enables (when $|c \rangle=|1 \rangle$) or disables (when $|c \rangle=|0 \rangle$) the accumulation operation (in the latter case the result is $|b \rangle$). Hence, the circuit uses a total of 3n + 1 qubits and its operation can be described as:

\begin{equation}
\mathit{\Phi}MAC_{a} (|c\rangle |x\rangle |b\rangle) = (|c\rangle |x\rangle |b+c \cdot ax\rangle)
\label{eq:FMAC}
\end{equation}

Taking into account the binary expansion of integer $x=(x_{n-1}, \ldots , x_{1}, x_{0})$, we can write the product $ax$ as:

\begin{equation}
ax=x_{0}a+x_{1}2a+ \cdots + x_{n-1}2^{n-1}a
\label{eq:expandedProd}
\end{equation}

\begin{figure} [!ht]
\centerline{\epsfig{file=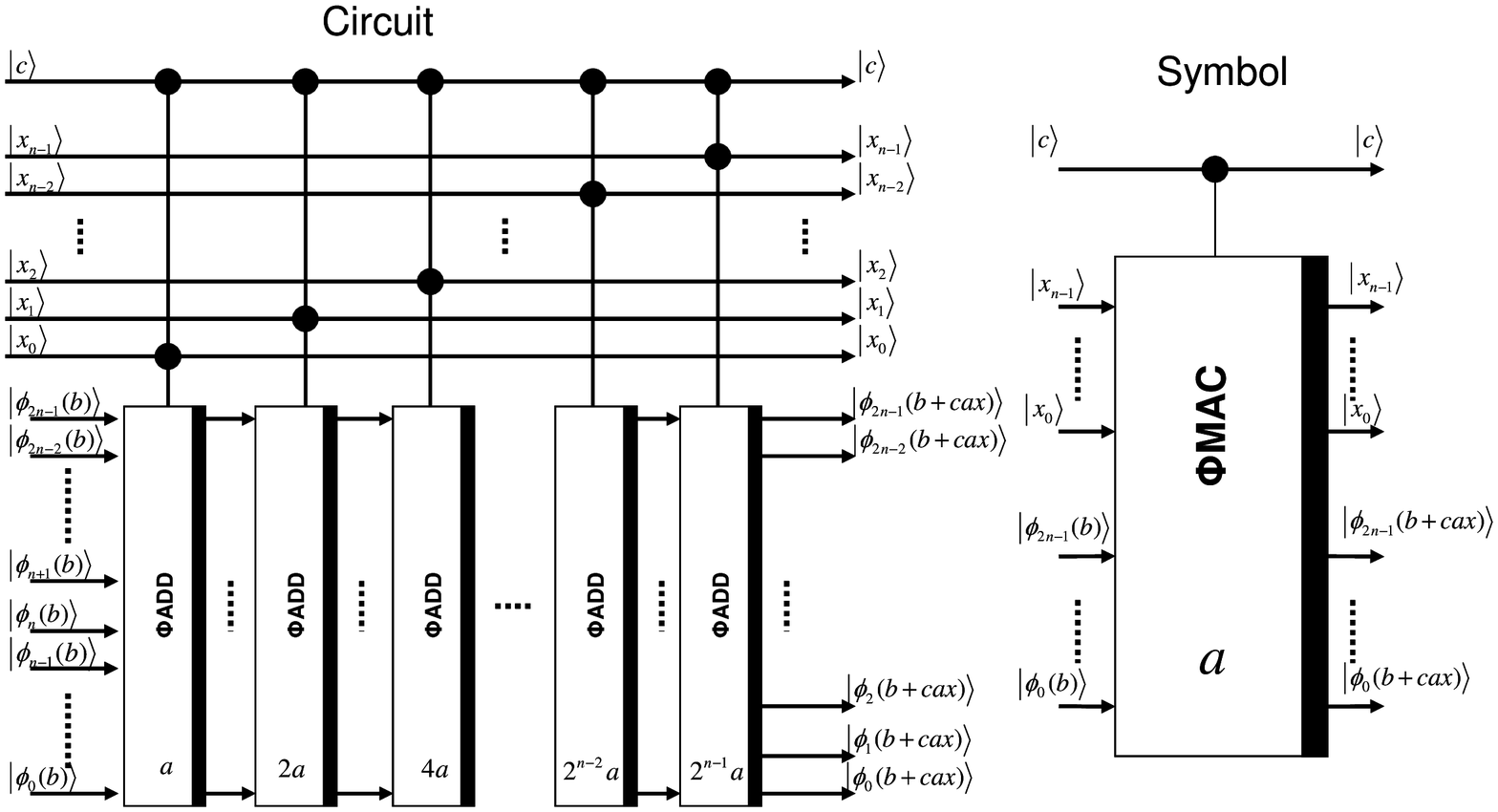, width=12cm}} 
\vspace*{13pt}
\fcaption{\label{fig:FMAC} Block level design of the multiplier/\- accumulator unit $\Phi$MAC  and its symbol. The basic blocks depicted here are the CC$\Phi$ADD units of Figure \ref{fig:CCFADD}. A detailed diagram of the above circuit is provided in Figure \ref{fig:FMACdetail}. }
\end{figure}

Therefore, the accumulation of the product $ax$ with $b$ can be achieved by the successive addition of $n$ constant integers $a,2a, \ldots ,2^{n-1}a$ each one being added conditionally on the qubit value $x_{j}, j = 0,1, \ldots, n-1$, respectively. Hence, the $\Phi$MAC circuit can be built as shown in Figure \ref{fig:FMAC}, assuming that the lowest $2n$ qubits (those that participate in the accumulation) are already transformed in the QFT representation by a previous QFT block. Actually, the circuit comprises a series of   CC$\Phi$ADD blocks described in the previous subsection, each one adding in succession the integers $a, 2a, \ldots , 2^{n-1}a$, and controlled by their two controlling qubits. The first controlling qubit $|c_{1} \rangle$ is common to all the CC$\Phi$ADD blocks and becomes the controlling bit for the $\Phi$MAC block. The second controlling qubit $|c_{2} \rangle$ of the $j^{th}$ adder CC$\Phi$ADD is the corresponding input qubit $|x_{j} \rangle, j=0, \ldots, n-1$   of the $\Phi$MAC block.

\begin{figure} [bt]
\centerline{\epsfig{file=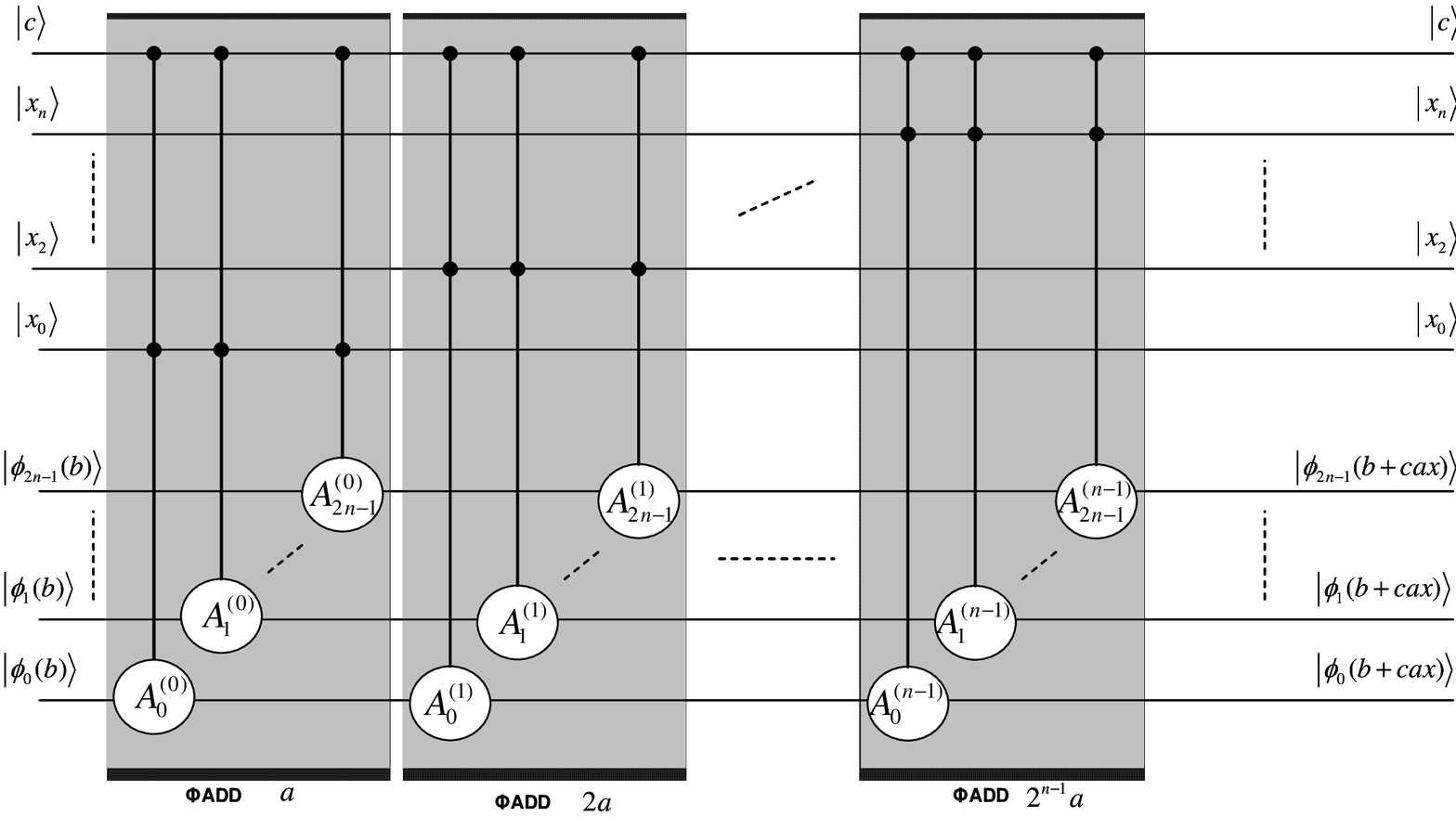, width=12cm}} 
\vspace*{13pt}
\fcaption{\label{fig:FMACdetail} Detailed design of the initial multiplier/accumulator $\Phi$MAC unit having depth $2n^{2}$. The depth improvement of this circuit is described in Section \ref{sec:FMAC} and the improved $\Phi$MAC is depicted in Figure \ref{fig:FMACD3x3}. }
\end{figure}

The detailed design of the $\Phi$MAC block that consists of doubly controlled rotation gates, is depicted in Figure \ref{fig:FMACdetail}. The $j^{th}$ CC$\Phi$ADD block adds to the $2n$ qubits that hold the accumulation result (in the QFT field) the integer $c^{(j)}=2^{j}a$, whose binary expansion is: 

\begin{equation}
c_{l}^{(j)}=a_{l-j}, j=0, \ldots, n-1
\label{eq:expandedConst}
\end{equation}

assuming that $a_{-1}, a_{-2}, \ldots, a_{-n}=0$.
The doubly controlled rotation gates $A_{l}^{(j)}$ in Figure \ref{fig:FMACdetail} affect the $l^{th}$ qubit of the accumulation register by using the following rotation matrix (if both the controlling qubits are in state $|1 \rangle$.

\begin{equation}
A_{l}^{(j)}=\prod\limits_{k=1}^{l+1}R_{k}^{c_{l+1-k}^{(j)}}=\prod\limits_{k=1}^{l+1}R_{k}^{a_{l+1-k-j}}= \left[
\begin{array}{cc}
1 & 0 \\
0 &
 e^{i2\pi\sum\limits_{k=1}^{l+1}\frac{a_{l+1-k-j}}{2^{k}}} 
\end{array} \right]
,
l=0, \ldots, 2n-1, j=0, \ldots, n-1
\label{eq:MultiplierAngles}
\end{equation}

Thus, constant number $a$ affects each adder of the $\Phi$MAC unit through Eq. \ref{eq:MultiplierAngles}, determining the rotation angle for each doubly controlled rotation gate.

\FloatBarrier

\section{Depth-Optimized Fourier Multiplier/Accumulator - {\textbf \textPhi}MAC}\label{sec:FMAC}
\noindent

 In this Section we describe one of the basic building blocks of the proposed designs for the modular exponentiation circuit. We modify the multiply/accumulate circuit ($\Phi$MAC) of Beauregard \cite{Beauregard} reducing its depth from $O(n^{2})$ to $O(n)$. 
 
The circuit of Figure \ref{fig:FMACdetail} is composed of $n$ CC$\Phi$ADD adders, each consisting of $2n$ doubly-controlled rotation gates with rotation matrix $A_{l}^{(j)}$  as described by Eq. \ref{eq:MultiplierAngles}, requiring a total of $2n^{2}$ such gates. All these gates have one common control qubit $|c \rangle$, which is the control qubit of the $\Phi$MAC unit. 

\begin{figure} [bt]
\centerline{\epsfig{file=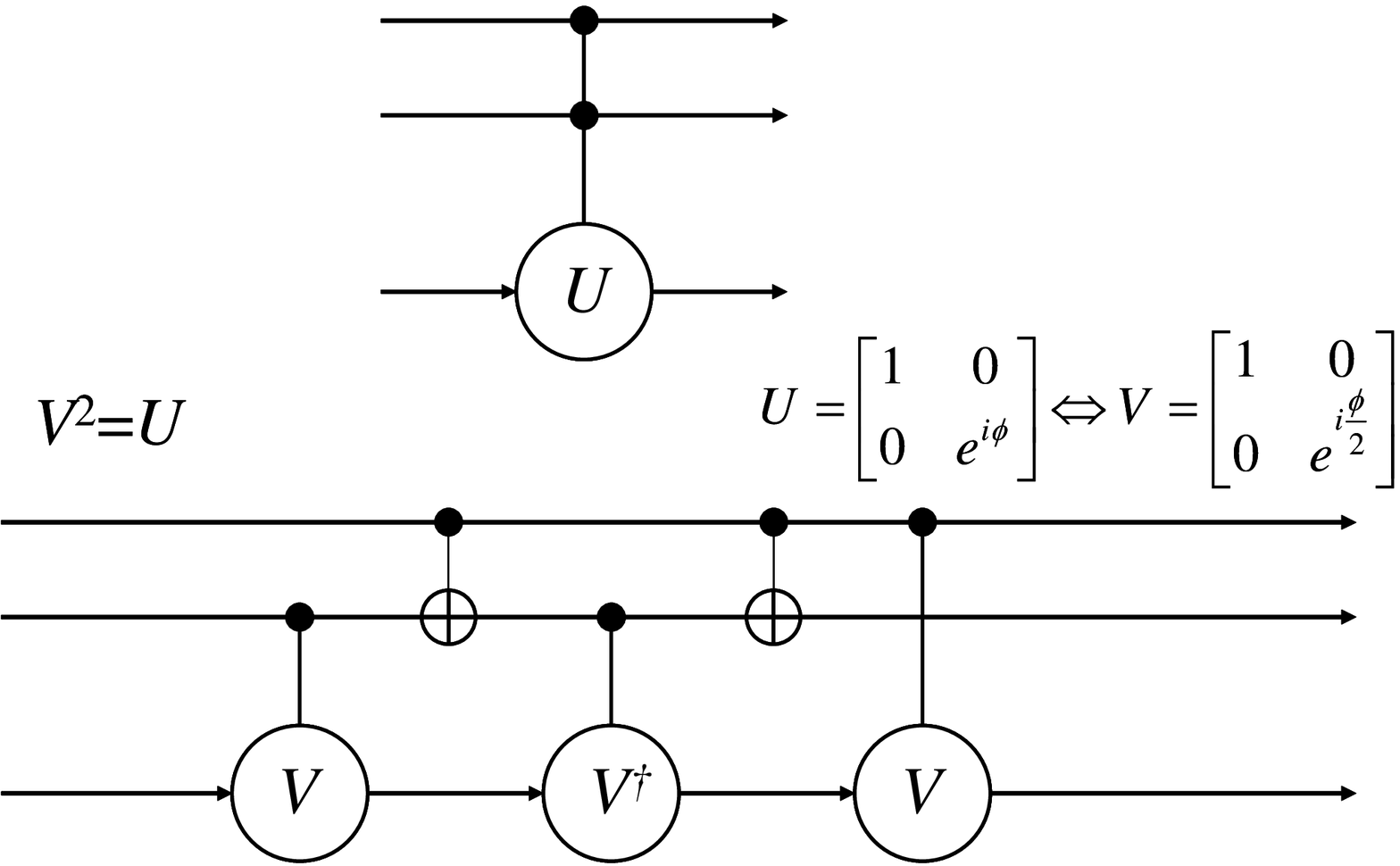, width=8.2cm}} 
\vspace*{13pt}
\fcaption{\label{fig:BarencoDecomposition} Doubly controlled three-qubit gate decomposition on a network of two-qubit gates. }
\end{figure}

\FloatBarrier

\begin{figure} [ht]
\centerline{\epsfig{file=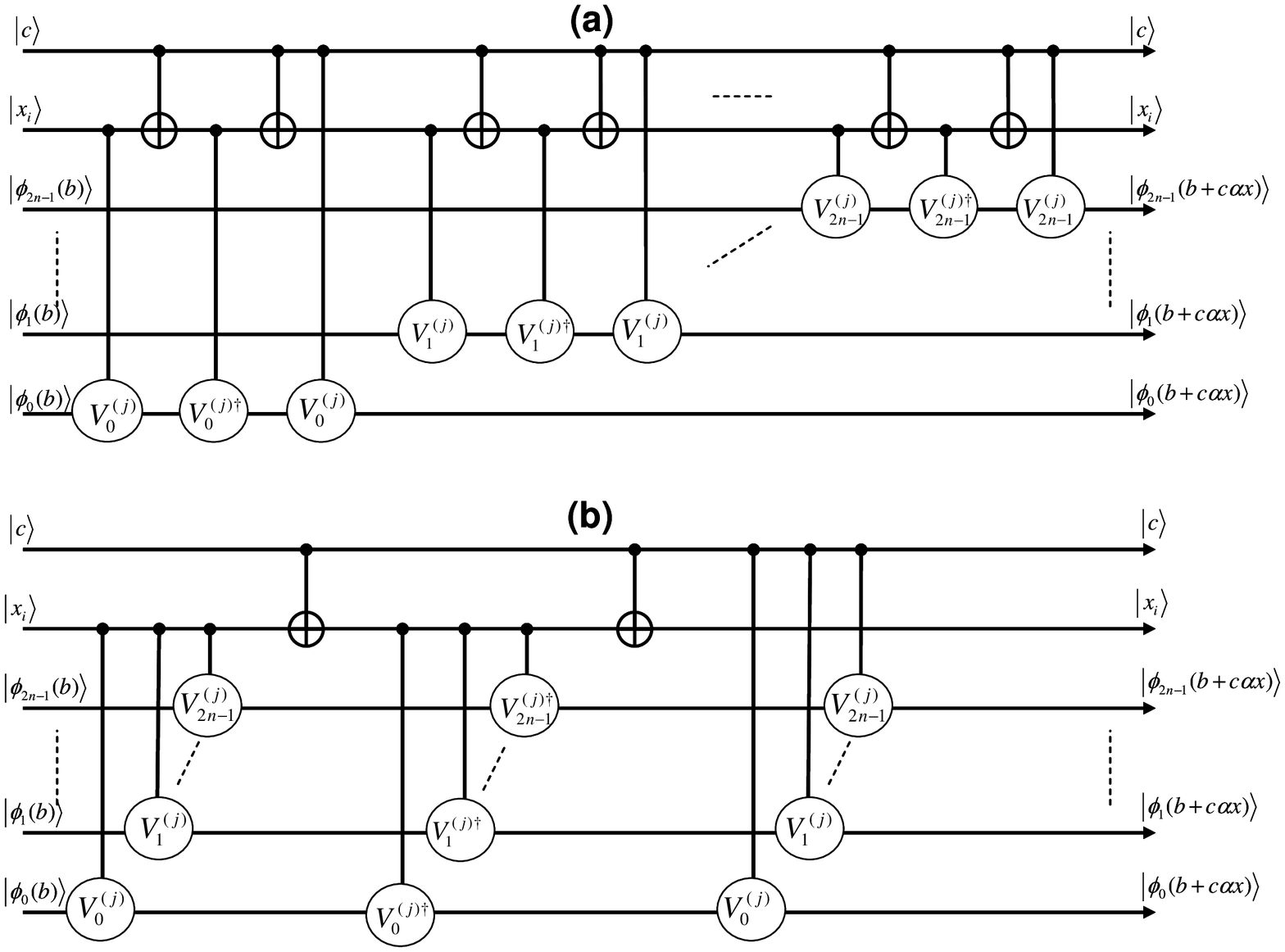, width=12cm}} 
\vspace*{13pt}
\fcaption{\label{fig:FMACDecomposition} (a) The $j^{th}$ $\Phi$ADD subcircuit of the $\Phi$MAC, (b) the rearrangement of the $j^{th}$ $\Phi$ADD subcircuit after exploiting the decomposition of Figure \ref{fig:BarencoDecomposition}. }
\end{figure}

\FloatBarrier

If we decompose each doubly controlled rotation gate into a network of single controlled gates \cite{Barenco1} as depicted in Figure \ref{fig:BarencoDecomposition}, we can re-arrange the rotation gates of the whole circuit so as to have a revised circuit with smaller depth. The matrices $V_{l}^{(j)}$ and $V_{l}^{(j)\dagger}$ of the rotation gates in Figure \ref{fig:FMACDecomposition} corresponding to the $\Phi$MAC matrices $A_{l}^{(j)}$, are given by the following equations:

\begin{equation}
V_{l}^{(j)}=\sqrt{A_{l}^{(j)}}= \left[
 \begin{array}{cc} 
1 & 0 \\
0 &  e^{i\pi\sum\limits_{k=1}^{l+1}\frac{a_{l+1-k-j}}{2^{k}}} 
\end{array} \right]
 ,l=0, \ldots ,2n-1, j=0, \ldots ,n-1
\label{eq:VAngles}
\end{equation}

\begin{equation}
V_{l}^{(j)\dagger}=\sqrt{A_{l}^{(j)}}^{\dagger}= \left[
\begin{array}{cc}
1 & 0 \\
0 &  e^{-i\pi\sum\limits_{k=1}^{l+1}\frac{a_{l+1-k-j}}{2^{k}}}
\end{array} \right]
 ,l=0, \ldots ,2n-1, j=0, \ldots ,n-1
\label{eq:VHAngles}
\end{equation}

A closer look of the subcircuit of Figure \ref{fig:FMACDecomposition}a that corresponds to the adder of constant $a2^{j}$ (the subcircuit corresponding to qubits $|c\rangle, |x_{j} \rangle, |\phi_{2n-1} \rangle, \ldots, |\phi_{1} \rangle,|\phi_{0} \rangle$) reveals that all the “first” $2n$ $V_{l}^{(j)}$ gates controlled by qubit $|x_{j} \rangle$ can be moved to the left of the subcircuit of Figure \ref{fig:FMACDecomposition}b, because they are all controlled by the same qubit $|x_{j} \rangle$ upon which a CNOT gate controlled by qubit $|c \rangle$ has acted an even number of times. This is equivalent to no CNOT gate acting. Similarly, all the “odd numbered” $2n$ CNOT gates that correspond to the decomposition of each $A_{l}^{(j)}$ gate can be replaced by exactly one CNOT gate affecting qubit $|x_{j} \rangle$ and controlled by qubit $|c \rangle$.

Next, all the  $V_{l}^{(j)\dagger}$ gates controlled by qubit $|x_{j} \rangle$ can be moved  exactly after the CNOT gate as shown in Figure \ref{fig:FMACDecomposition}b, because their control is done by the qubit $|x_{j} \rangle$, upon which a CNOT gate controlled by qubit $|c \rangle$ has acted an odd number times. This is equivalent to only one CNOT gate. Also, the “even numbered” group of $2n$ CNOT gates corresponding to the decomposition of each $A_{l}^{(j)}$ gate are merged to a simple CNOT gate exactly after the grouping of the controllable $V_{l}^{(j)\dagger}$ gates. Finally, using the same arguments as before we can merge the “last” $2n$ $V_{l}^{(j)}$ gates at the right of the subcircuit of Figure \ref{fig:FMACDecomposition}b.

After these transformations, we can combine the $n$ quantum adders CC$\Phi$ADD in a highly parallel circuit as depicted in Figure 12 for the case $n = 3$. Figure 12 refers to the case of multiplying a three bit integer constant $a$ with a three qubits quantum integer $x$ and accumulating the resulting product into a six qubits quantum register. This circuit is built by successively connecting $n$ CC$\Phi$ADD blocks and exploiting the fact that all the controllable rotation gates commute. Furthermore, the last $V_{l}^{(j)}$ gates acting upon qubit $|\phi_{l} \rangle$ can be merged (as long as they are all controlled by qubit $|c \rangle$  ) to a single controlled gate $W_{l}$, with rotation matrix:

\begin{equation}
W_{l}=\prod\limits_{j=1}^{n-1}V_{l}^{(j)},l=0, \ldots ,2n-1
\label{eq:WAngles}
\end{equation}

\begin{figure}[H]
\centering
\includegraphics[totalheight=0.48\textheight, angle=90]{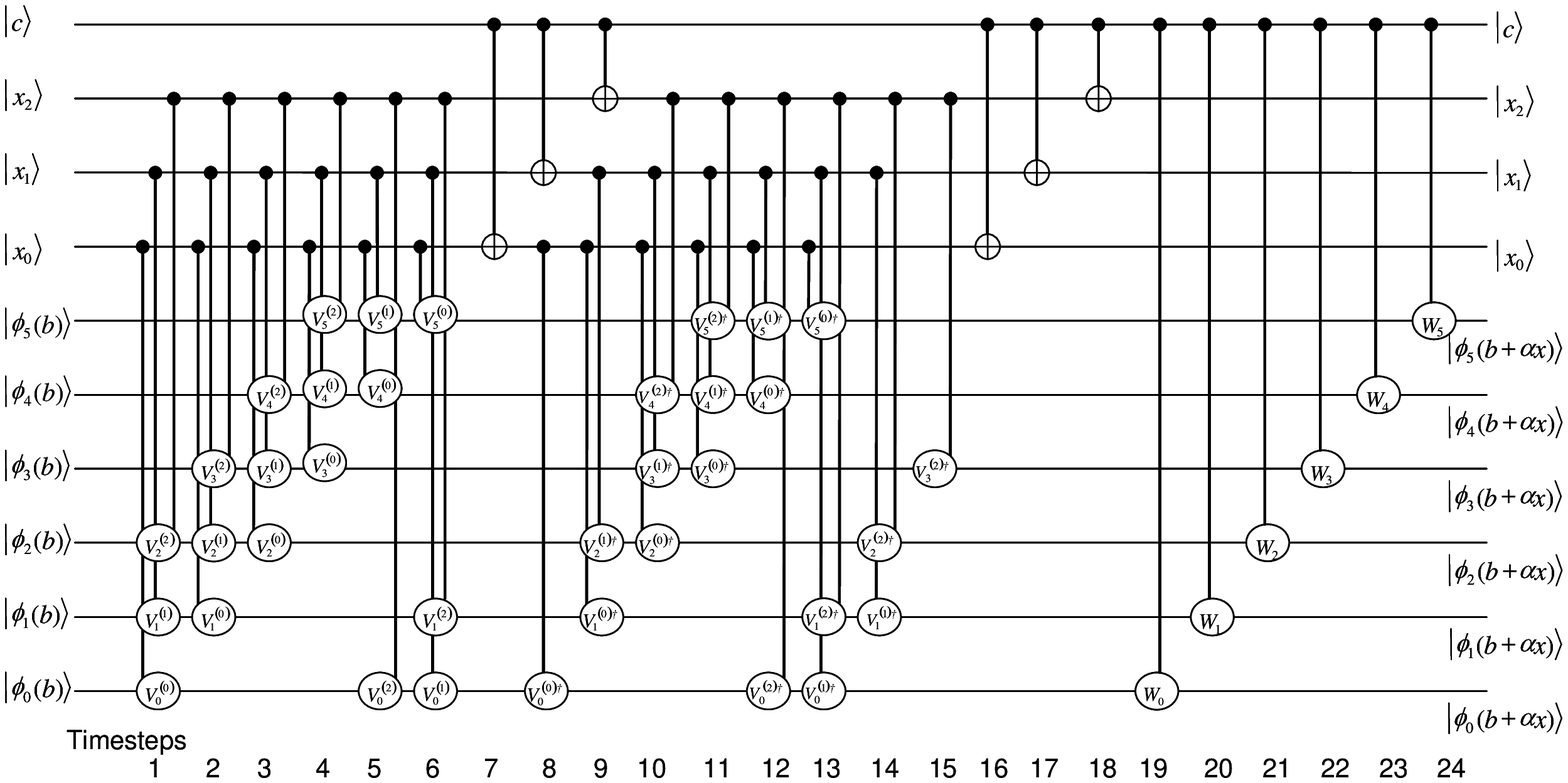} %
\vspace*{13pt}
\fcaption{\label{fig:FMACD3x3} Fully decomposed and rearranged $\Phi$MAC unit with linear depth of $8n$ for the case $n=3$. In this case it requires $8\cdot3$=24 timesteps as shown in the figure. The rotation gates angles are determined by the constant $a$ (see Eq. \ref{eq:VAngles}, \ref{eq:VHAngles} and \ref{eq:WAngles}).}
\end{figure}

A circuit depth analysis for the $\Phi$MAC unit of Figure \ref{fig:FMACD3x3} shows that if the two-qubit gates acted upon and controlled by different qubits operate in parallel, then for the “first” $V_{l}^{(j)}$ gates a total of $2n$ computation steps are required, for the “odd group’ CNOT gates and the $V_{l}^{(j)\dagger}$, $3n$ more computations steps are require, for the “even group” CNOT gates $n$ more computation steps are required, and for the $W_{l}$ gates, $2n$ more computation steps are required. Consequently, the total required computational steps required for the proposed implementation of the multiplier/accumulator unit is linear in size and has a depth of $8n$. Furthermore, the proposed circuit uses almost exclusively two-qubit rotation gates, instead of three-qubit gates, making it more suitable for physical realizations \cite{DiVincenzo,Lanyon,Ralph,Monz,Fedorov}. Fault tolerance aspects of the circuit depth are not taken into account. This point is separately addressed in Section \ref{sec:Analysis}.


\section{QFT Divider by constant - GM{\textbf\textPhi}DIV}\label{sec:GMFDIV}
\noindent
The proposed modular exponentiation design does not use a modular adder to construct the multiplier/accumulator unit but it is rather based on simple adders. For this reason, we are forced to implement the modular operation after the multiplication by incorporating a divider module. Dividers are the most complex elementary arithmetic operation circuits in terms of computation time, but for our Shor's algorithm circuit design we can again take advantage of the fact that only divisions by the integer to be factored, $N$, are required. Integer $N$ is constant throughout each quantum iteration of Shor's algorithm and thus a simpler division module suffices.

There are a few quantum dividers known, among them there are some \cite{Kaye,Amento}  which are suitable for multiplicative inversion in the Galois field GF($2^{m}$) with depth of $O(n^{3})$ and $O(nlog_{2}n)$, respectively. Another divider suitable for integer division appears in \cite{Khospour} and is based in QFT with a depth of $O(n^{3})$. This divider receives two quantum integers and gives the quotient and the remainder. If one tries to convert it to a divider by constant then its depth can be reduced to $O(n^{2})$.
 
Various algorithms for classical division of an integer by constant have been appeared in the literature, such as those in \cite{Granlund} and \cite{Moller}. In this section we describe a quantum version of an algorithm proposed by Granlund and Montgomery in \cite{Granlund}. This algorithm divides a $2n$ bits integer by an $n$ bits constant integer and generates an $n$ bits quotient and an $n$ bits remainder, subject to the constraint that the quotient is less than $2n$. The algorithm can be easily modified to an unconstrained algorithm that divides an $n$ bits integer by an $n$ bits constant integer by simply zeroing the upper $n$ bits of the dividend. The algorithm in \cite{Granlund} utilizes multiplications, additions, logical operations such as shifting and bit selections. It has a constant time complexity. Table \ref{table:algo} shows the algorithm in pseudocode format as presented in \cite{Granlund}.

\vspace*{4pt}   
\begin{table}[!b]
\tcaption{Granlund-Montgomery division-by-constant algorithm \cite{Granlund}. Comments, enclosed in \mbox{\texttt{/*  */}},  in some of the lines show equivalent arithmetic operations.}
\centerline{\footnotesize\smalllineskip
\begin{tabular}{r l c l c }\\
\hline
    & \texttt{udword }$z$\texttt{;} & \texttt{/*} & Dividend (input)     & \texttt{*/} \\
    & \texttt{uword }$q$\texttt{;} & \texttt{/*} & Quotient (output)    & \texttt{*/} \\
    & \texttt{udword }$r$\texttt{;} & \texttt{/*} & Remainder (output)   & \texttt{*/} \\
    & \texttt{const uword }$d$\texttt{;} & \texttt{/*} & Divisor (constant) & \texttt{*/} 
\\
& \multicolumn{4}{l}{\texttt{/*}  Initialization (given uword  $d$, where $0<d<2^{n}$) \texttt{*/}} \\
\texttt{1:}    & \texttt{uword }$l=1+$\texttt{floor}$(log_{2}d)$\texttt{;} & \texttt{/*} & $2^{l-1} \leq d<2^{l}$ & \texttt{*/} \\
\texttt{2:}    & \texttt{uword }$m'=$ \texttt{ floor}$(2^{n}·(2^{l} -d ) -1/d)$\texttt{;} & \texttt{/*} & $m'=$\texttt{floor}$((2^{n+1}-1)/d)-2^{n}$ & \texttt{*/} \\
\texttt{3:}    & \texttt{uword }$d_{norm}=$\texttt{ SLL}$(d,n-l)$\texttt{;} & \texttt{/*} & Normalized divisor $d\cdot 2^{n-l}$ & \texttt{*/} \\
& \multicolumn{4}{l}{\texttt{/*}  Start of main procedure  \texttt{*/}} \\
\texttt{4:}    & \texttt{uword }$n_{2}=$ \texttt{ SLL}$($\texttt{HIGH}$(z),n-l)$\texttt{+SRL}$($\texttt{LOW}$(z),l)$\texttt{;} & & &  \\
\texttt{5:}    & \texttt{uword }$n_{10}=$\texttt{ SLL}$($\texttt{LOW}$(z),n-l)$\texttt{;} & & &  \\
\texttt{6:}    & \texttt{sword }$n_{1}=-$\texttt{XSIGN}$(n_{10})$\texttt{;} & & &  \\
\texttt{7:}    & \texttt{uword }$n_{adj}=n_{10}+$\texttt{AND}$(-n_{1},d_{norm}-2^{n})$\texttt{;} &\texttt{/*} & $n_{adj}=n_{10}+ n_{1} \cdot (d_{norm}-2^{n}) $ & \texttt{*/} \\
\texttt{8:}    & \texttt{uword }$q_{1}=n_{2}+$\texttt{HIGH}$(m'\cdot(n_{2}+n_{1})+n_{adj})$\texttt{;} & & &  \\
\texttt{9:}    & \texttt{sdword }$dr=z-2^{n}\cdot d+(2^{n}-1-q_{1})\cdot d$ \texttt{;} &\texttt{/*} & $ dr=z-q_{1}\cdot d - d, -d \leq dr < d $ & \texttt{*/} \\
\texttt{10:}    & $q=$ \texttt{HIGH}$(dr)-(2^{n}-1-q_{1})+2^{n}$ \texttt{;} &\texttt{/*} &  Add 1 to quotient if $dr\geq 0 $ & \texttt{*/} \\
\texttt{11:}    & $r=$ \texttt{LOW}$(dr)+$ \texttt{AND} $(d-2^{n},$ \texttt{HIGH} $(dr))$  \texttt{;} &\texttt{/*} & Add $d$ to remainder if $dr<0$                                    & \texttt{*/} \\
\hline\\
\end{tabular}}
\label{table:algo}
\end{table}

\FloatBarrier

This division algorithm takes as input the unsigned integer $z$ (dividend) and divides it by the constant unsigned integer $d$ (divisor) giving as results the quotient $q$ and the remainder $r$.  The three computation steps of the initialization (lines 1-3 of Table \ref{table:algo}) are to be done once, as they depend only on the constant divider. These initialization steps are “hardwired” in the quantum version of the algorithm and reflect a particular divisor which is the number to be factored. The computation steps of the main division procedure (lines 4-11) are executed whenever a new dividend must be divided by the constant divisor, that is at each iteration of the quantum part of Shor's algorithm for each new random number $a$. An explanation of the meaning of the variables and data types of the algorithm along with the various logical operations follows in Table \ref{table:definitions}. Also shown is the required number of ancillae for each operation.

\vspace*{4pt}   
\begin{table}[!b]
\tcaption{Explanation of the various logical operations and data types used in the classical version of the division by constant algorithm.}
\centerline{\footnotesize\smalllineskip
\begin{tabular}{|l|l|c|c|c|}
\hline
\multicolumn{1}{|c|}{Operation}  & \multicolumn{1}{|c|}{Meaning} & Input  & Output  & Ancilla  \\
\multicolumn{1}{|c|}{Type} &  &  (\# of bits) &  (\# of bits) & (\# of bits)  \\ \hline
\texttt{SLL} $(x,i)$ & Logical left shift of $x$ by $i$ bits & $n$ & $n$ & $n$ \\
\texttt{SRA} $(x,i)$ & Arithmetic right shift of $x$ by $i$ bits & $n$ & $n$ & $n$ \\
\texttt{SRL} $(x,i)$ & Logical right shift of $x$ by $i$ bits & $n$ & $n$ & $n$ \\
\texttt{XSIGN} $(x)$ & -1 if $x<0$, 0 if $x\geq 0$ & $n$ & $1$ & $1$ \\
\texttt{AND} $(x,y)$ & Bitwise logical AND of $x$ and $y$ & $n$ & $n$ & $0$ \\
\texttt{HIGH} $(x)$ & Upper half of integer  $x$  & $2n$ & $n$ & 0 or $n$ \\
\texttt{LOW} $(x)$ & Lower half of integer  $x$  & $2n$ & $n$ & 0  \\
\texttt{uword} & $n$ bits unsigned integer  &   &   &    \\
\texttt{sword}  & $n$ bits signed integer &   &   &    \\
\texttt{udword}  & $2n$ bits unsigned integer  &   &   &    \\
\texttt{sdword}  & $2n$ bits signed integer  &   &   &    \\
\hline
\end{tabular}}
\label{table:definitions}
\end{table}

The shift operations above (\texttt{SLL}, \texttt{SRA}, \texttt{SRA}) can be easily implemented reversibly (without any bits discarding) with the help of ancilla register initialized with value 0. As shown in lines 4 and 5 of the algorithm, the result of the shifting is to be added  to the valued 0 or added one after the other to the value 0, thus we can just select the desired bits to drive the input of the generic $\Phi$ADD unit of Figure \ref{fig:GFADD}, while the QFT transformed input of the $\Phi$ADD unit is the ancilla. The original input to be shifted will remain intact in the initial register. Of course the ancilla register has to be set back to the zero value somehow at a later time. 

The\texttt{ XSIGN} operation is simply a copying of the most significant bit (the sign bit) of a register holding a signed integer, to an ancilla qubit. Similarly, this ancilla has to be zeroed later.

The \texttt{AND} operations are implemented as the arithmetic operations shown in the comments of lines 7 and 11 of the algorithm of Table \ref{table:algo}, that is additions on condition of the value of $n_{1}$ and $dr$, respectively.

The \texttt{LOW} operations in lines 4 and 5 of the algorithm are just the selection of the relevant bits to be “shifted” and then added by the QFT adders as described above. The third \texttt{LOW} in line 11 of algorithm is done implicitly as an addition shown in the comment of line 11. 

The \texttt{HIGH} operation in line 4 is implemented with the help of an ancilla register, initially in value 0. The upper half bits are copied via CNOT gates to the ancilla register and later this ancilla must be reset to the zero value. The \texttt{HIGH} operation in line 11 is accomplished similarly to the \texttt{LOW} as the addition shown in the comment.

A quantum division by constant circuit implementing the algorithm of \cite{Granlund}  is depicted in the two diagrams of Figures \ref{fig:GMFDIVa} and \ref{fig:GMFDIVb} (subsequently referred as Figures \ref{fig:GMFDIVa}-\ref{fig:GMFDIVb}). A quantum circuit for this algorithm needs a few ancilla qubits because of the intermediate variables used in its classical counterpart, like $m'$, $d_{norm}$, $n_{2}$, $n_{10}$, $n_{1}$, $n_{adj}$, $q_{1}$, $dr$ and these ancilla qubits must be zeroed again (assuming initial zero state) at the end of the computation, since we want to reuse them for subsequent computations. Figures \ref{fig:GMFDIVa}-\ref{fig:GMFDIVb} show a quantum circuit that implements the algorithm of Table \ref{table:algo} for the case $n = 4$ and for a constant divisor $d = 5$. The extension for other sizes of $n$ and different constant divisors $d$ is straightforward; we refer to such division circuits as GM$\Phi$DIV.

\subsection{Building block and registers of the quantum divider.}
\noindent
A generic GM$\Phi$DIV circuit will have a total of $7n + 1$ qubits, $5n + 1$ of which are the ancilla qubits. The GM$\Phi$DIV unit uses the following blocks and their inverses (inverses are noted with the same symbol with the thick bar on the left instead of the right): 
\begin{itemize}
\item
QFT for computing the quantum Fourier transform.

\item
Three types of adders $\Phi$ADD, that is adder with quantum integer, adder with constant and controlled adder with constant. Their inverses are simply the reverse circuit (signal flow from output to input) with the angles of rotation gates having the opposite sign.

\item
Multiplier/accumulator $\Phi$MAC and its inverse which is simply the reverse circuit with opposite sign angles in its rotation gates.

\item
CNOT, $X$ (NOT) gates and SWAP gates (implied by the rerouting of the registers around the third $\Phi$MAC unit).
\end{itemize}

The input qubits of the GM$\Phi$DIV circuit are grouped in a   qubits register (Reg0:Reg1), five  qubits registers (Reg2,$\ldots$,Reg6), most of which are ancillae, and a single ancilla qubit (Aqbit) as shown in Figures \ref{fig:GMFDIVa}-\ref{fig:GMFDIVb}. We give below a description of the purpose of each register:

\begin{itemize}
\item
Reg0:Reg1: We distinguish two cases:
	\begin{alphlist}
	\item
	In the generic divider case, where we divide an $n$ qubits number by an $n$ bits constant, this register initially contains the dividend $z$ in the qubits of its lower half, while the upper half is initially set to zero. This way we conform to the constraint that the quotient must be less than $2n$. At the end of the computation, this register will contain the remainder $r$ in its lower half and zero in its upper half. The upper half acts essentially as an ancilla register. 
	\item
	In the special case where we divide a $2n$ qubits number by an $n$ bits constant, under the restriction that the quotient is known to be less than $2n$, both Reg0 and Reg1 will contain the upper and lower part of dividend, respectively. 
	Since both cases of division differ only in the permitted type of initialization in registers Reg0 and Reg1, otherwise they have the same circuit network, we distinguish these cases by different symbols as shown later. 
\end{alphlist}

\item
Reg2: This register contains the quotient $q$ at the end of the computation. It is initialized in the zero state.

\item
Reg3:  Ancilla register holding the intermediate variable $q_{1}$. Initialized and end up in zero state.

\item
Reg4: Ancilla register used to successively hold the values $n_{2}$, $n_{2}+n_{1}$, $n_{2}$, 0, $n_{1}$, 0. Initialized and end up in zero state. 

\item
Reg5: Ancilla register used to hold the value \texttt{HIGH}$(m'(n_{2}+n_{1})+n_{adj})$. Initialized and end up in zero state.

\item
Reg6: Ancilla register used to successively hold the values $n_{10}$, $n_{adj}$, \texttt{LOW}$(m'(n_{2}+n_{1})+n_{adj})$.

\item
Aqbit: Ancilla qubit used to hold the sign $n_{1}$. Initialized and end up in zero state.

\end{itemize}

Next a brief description of the whole circuit is given. Part of the circuit is dedicated to forward computations, while another part is dedicated to “restoring” the ancilla qubits back to the zero state, so as the whole circuit is reversible. The latter part stands out as the gray shaded area. We begin describing the forward computations. For simplicity we refer to the values of the registers as integers regardless of being integers in the computational basis or being their respective values in the quantum Fourier transform domain, in other words we ignore in the description the various QFT blocks present in the register buses. In essence QFT and inverse QFT blocks are needed at the buses before and after, respectively, each arithmetic block that process integers in the Fourier domain. These are: the bus of one of the integers of every kind of adder ( $\Phi$ADD, C$\Phi$ADD ) and the accumulator bus of each $\Phi$MAC block. Whenever two adders are connected one after the other in order to successively add different values to a quantum integer, there is no need to transform from the QFT domain back to the computational basis and then again perform the forward transform. The schematic diagram is adequate to distinguish when a bus has an integer in the computational basis or in the Fourier domain.

\subsection{Forward computations of the quantum divider.}

\noindent
\underline{Initialization steps (Lines 1, 2, 3, of the Algorithm)}

\begin{figure}[H]
\centering
\includegraphics[totalheight=0.65\textheight, angle=270]{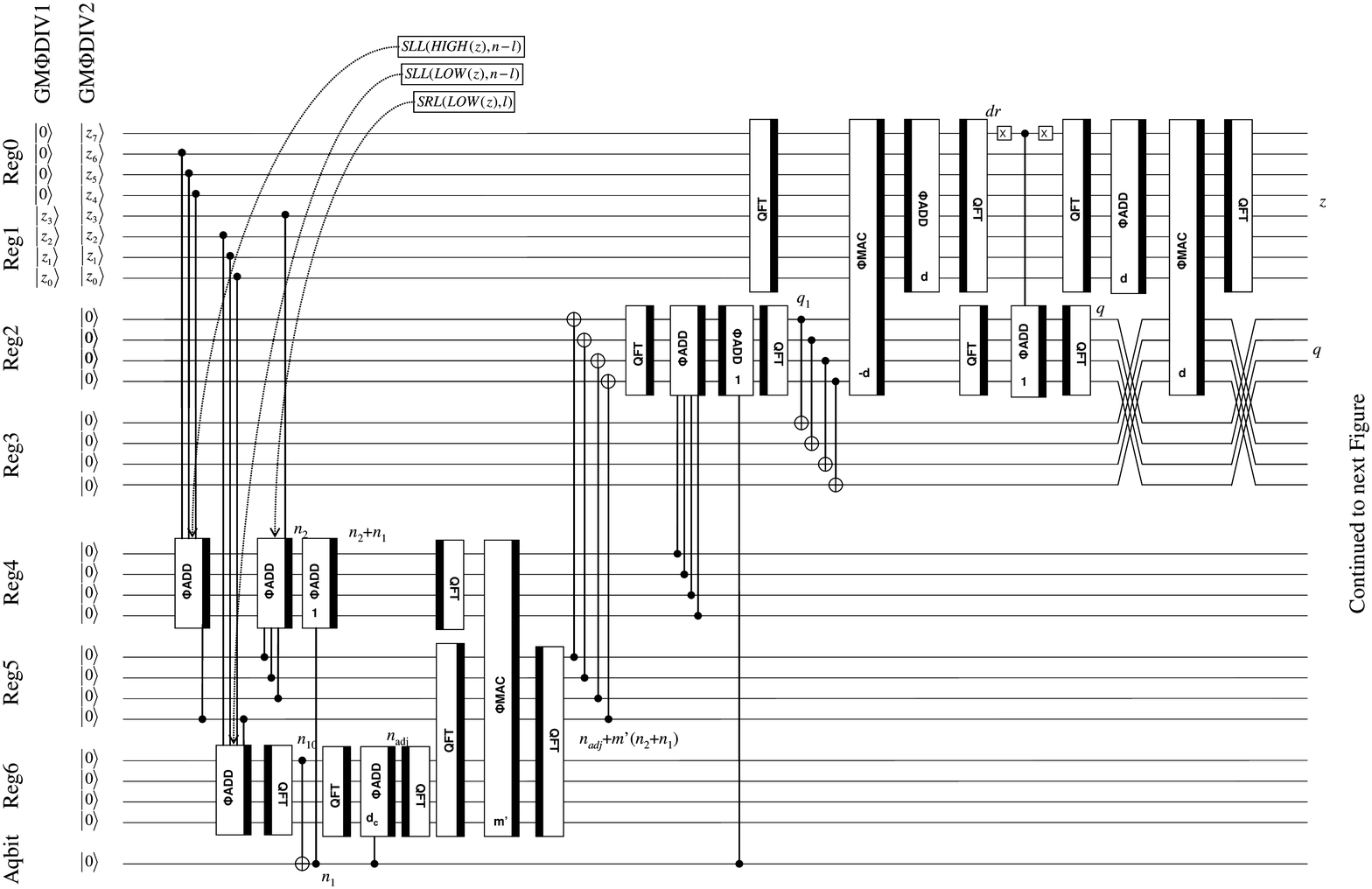} %
\vspace*{13pt}
\fcaption{\label{fig:GMFDIVa} The GMΦ$\Phi$DIV circuit (first part) for 8 or 4 qubits dividend and constant divisor $d=$5. Intermediate variables are shown at places where they have been computed (in computational basis or QFT transformed).}
\end{figure}

\begin{figure}[H]
\centering
\includegraphics[totalheight=0.65\textheight, angle=270]{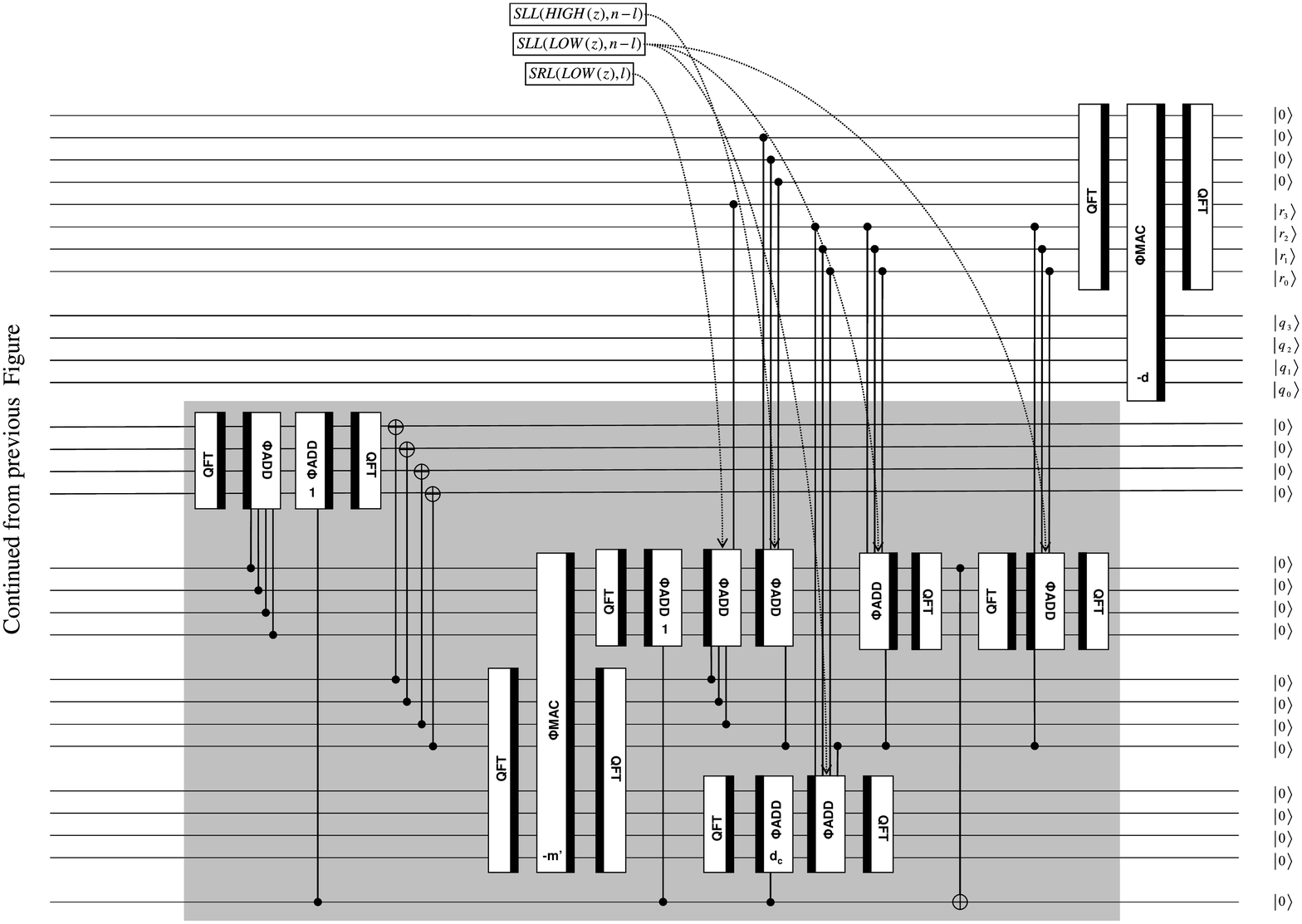} %
\vspace*{13pt}
\fcaption{\label{fig:GMFDIVb} The GM$\Phi$DIV circuit (second part) for 8 or 4 qubits dividend and constant divisor $d=5$. Shaded areas indicate computations for resetting the ancilla qubits.}
\end{figure}

\noindent
As noted before, the circuit of Figures \ref{fig:GMFDIVa}-\ref{fig:GMFDIVb} refers to the case of $d=5$. The initialization steps in lines 1, 2 and 3 of the algorithm of Table \ref{table:algo} result in the following values of $l=3$, $d_{norm}=10$ (and $d_{norm}-2^{n}=-6$) and $m'=9$. This initialization is computed “offline”. These values are to be “hardwired” in the circuit, i.e. $m'=9$ is hardwired as the constant parameter of the first $\Phi$MAC unit, $d_{c}=d_{norm}-2^{n}=-6$ is hardwired as the constant parameter in some of the C$\Phi$ADD units and $l=3$ is hardwired in the logical shift section. The logical shift sections are indicated by dashed arrows [pointing from the logical operation to the adders performing the shift. 

\noindent
\underline{Computation of $n_{2}$, $n_{10}$, $n_{1}$. (Lines 4, 5, 6 of Algorithm)}

\noindent
The first operation in the diagram is to compute the value of $\texttt{SLL}(\texttt{HIGH}(z),n-l)=(z_{6}z_{5}z_{4}0)_{2}$ to be added to Reg4. Indeed, we select these three qubits of   from Reg0 along with a zero qubit from ancilla Reg5 and add them to Reg4 through the use of an $\Phi$ADD unit. (This operation is to be done only in the special case (b) of the divider mentioned above where both halves of the dividend may contain non-zero values. In the general case where the upper half contains zero values this operation becomes gratuitous). The next adder affecting Reg4 is the one that selects three zero qubits from Reg5 as most significant qubits and the qubit $z_{3}$ to add to the Reg4, that is to add $\texttt{SRL}(\texttt{LOW}$(z),l)$=\texttt{SRL}(\texttt{LOW}(z),3)=(000z_{3})$. This way we have computed in Reg4 the quantity $n_{2}$. In the same manner we compute in Reg6 the value of $n_{10}$. Having computed $n_{10}$, it is straightforward to compute the sign $n_{1}$ in the Aqbit by using  a CNOT gate controlled by the most significant qubit of Reg6 and targeting the Aqbit.  

\noindent
\underline{Computation of $_{nadj}$, $q_{1}$ (Lines 7,8 of Algorithm)}

\noindent
Now, we add the constant $d_{c}$ (that is $d_{norm}-2^{n}$) to Reg6 (having already the value  $n_{10}$) conditioned on the value of $n_{1}$, thus forming the quantity $n_{adj}$. Also, we add $n_{1}$ to Reg4, forming the value $n_{2}+n_{1}$. Now the first $\Phi$MAC($m'$) unit has at its accumulator input (high qubits at Reg5, low qubits at Reg6) the value  $n_{adj}$ and has at its multiplicand input (Reg4) the value $n_{2}+n_{1}$. Thus, the $\Phi$MAC outputs have Reg4=$n_{2}+n_{1}$ and (Reg5:Reg6)=$m'(n_{2}+n_{1})+n_{adj}$. By copying with CNOT gates the content of Reg5  to Reg2 we have at Reg2 the value $\texttt{HIGH}(m'(n_{2}+n_{1})+n_{adj})$. Then we can add to Reg2 the value of Reg4, which is still $n_{2}+n_{1}$ and then we subtract $n_{1}$ leaving as end result the desired $q_{1}=n_{2}+\texttt{HIGH}(m'(n_{2}+n_{1})+n_{adj})$.  

\noindent
\underline{Computation of $d_{r}$ (Line 9 of Algorithm)}

\noindent
The second $\Phi$MAC(-$d$) unit has at the accumulator input (Reg0:Reg1) the dividend $z$ and at the multiplicand input (Reg2) the value $q_{1}$, forcing the output (Reg0:Reg1) to be $z-d\cdot q_{1}$ and after subtracting the constant $d$ becomes  $d_{r}=z-d \cdot q_{1}-d$. 

\noindent
\underline{Computation of the quotient $q$ (Line 10 of Algorithm)}

\noindent
Now that we have computed the $d_{r}$ value we are ready to proceed to the last steps of the algorithm of Table 1\ref{table:algo} doing a sign check to the quantity $d_{r}$ as these steps suggest and this is equivalent to checking the most significant qubit of Reg0:Reg1. For this reason we add the integer 1 to the Reg2 conditionally on the inverted most significant qubit of Reg0:Reg1. This way we have formed at the Reg2 the quotient $q$, because if $d_{r}\geq 0$  then its inverted most significant qubit will be 1 thus adding the value 1 to $q_{1}$, otherwise it adds nothing. 

\noindent
\underline{Computation of the remainder $r$ (Line 11 of Algorithm)}

\noindent
Meanwhile $q_{1}$ has been copied to Reg3 by CNOT gates and becomes the multiplicand input of the third $\Phi$MAC($d$)  unit. Its accumulator register Reg0:Reg1 has become again $z-d \cdot q_{1}$ after the addition of $d$, and the end result for the accumulator register after the third $\Phi$MAC($d$) is to restore its initial value of the dividend $z$. The last $\Phi$MAC($-d$) unit acts on this register again, subtracting the product $q\cdot d$, giving thus the remainder $r$ at its lower half while its upper half becomes zero, thus completing the forward computations. Note that in both the generic case (dividend of $n$ qubits) and in the special case (dividend of $2n$ qubits subject to the constraint that the quotient is known to be less than $2^{n}$) the upper half (Reg0) becomes zero.

\subsection{Ancilla Restoration.}
\noindent
It remains now to show the computations that reset the ancilla qubits. The ancilla qubits to be reset are those in registers Reg3, Reg4, Reg5, Reg6 and the single qubit Aqbit. 

\begin{itemize}
\item
Reg3: The first reset occurs to Reg3 which contains $q_{1}$. This is accomplished if we subtract from it the quantity $n_{2}+n_{1}$ (stored in Reg4) and adding $n_{1}$ (stored in Aqbit) leaving a value of $\texttt{HIGH}(m'(n_{2}+n_{1})+n_{adj})$. But this value is already stored in Reg5 and a “qubitwise” CNOT operation from Reg5 to Reg3 effectively resets Reg3. 

\item
Reg5: Then we reset Reg5 through the usage of an $\Phi$MAC($-m'$), that is we add to the accumulator registers (Reg5:Reg6)  containing $m'(n_{2}+n_{1})+n_{adj}$ the quantity $-m'(n_{2}+n_{1}$) leaving the result $n_{adj}$. But $n_{adj}$ is an \texttt{uword} and consequently the upper register (Reg5) becomes zero. 

\item
Reg4: Next we reset Reg4 by subtracting from it the quantities $n_{2}$ and $n_{1}$. Quantity $n_{10}$ is formed again easily from Reg0:Reg1 which now contains again the dividend $z$, by using the same method of shifting and additions we used in the forward computations. The value of $n_{10}$ is needed to reset the Aqbit as described below. Then by substracting $n_{10}$ we end up in the zero value in Reg4.

\item
Reg6: This register, which contains $n_{adj}$, is reset by subtracting the constant $d_{c}$ conditioned on $n_{1}$ and subtracting $n_{10}$. 

\item
Aqbit: To reset qubit Aqbit we use the value $n_{10}$ formed in Reg4 at the step just before we reset it for the second time and use its most significant qubit to CNOT the Aqbit.
\end{itemize}

\begin{figure} [htb]
\centerline{\epsfig{file=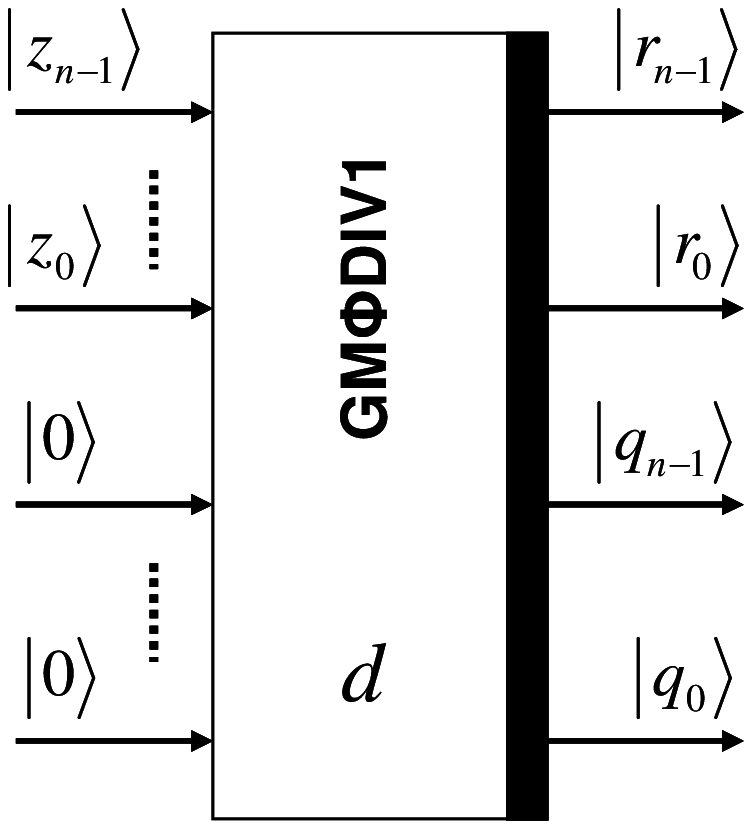, width=4cm}} 
\vspace*{13pt}
\fcaption{\label{fig:GMFDIV1} The symbol of the GM$\Phi$DIV1. It receives an $n$ qubits dividend and the underlying circuit uses $7n+1$ qubits, including $5n+1$ ancilla qubits which are not shown.}
\end{figure}

Note that the QFT divider proposed doesn't include any controlling qubit, but this is not required for the construction of the modular multiplier as will be shown. A symbol for the GM$\Phi$DIV for the generic case, that is dividend and divisor of $n$  bits wide, is shown in Figure \ref{fig:GMFDIV1} with the name GM$\Phi$DIV1. This symbol shows only the qubits used for input (dividend with the upper qubits having initial value of zero) and outputs (quotient and remainder), leaving hidden the other ancilla qubits.  

Regarding the GM$\Phi$DIV circuit, a thorough depth analysis leads to a depth of $244n-8$. In this study we have take in account the following facts. Each QFT unit has a depth of $2n-1$ (through parallelization of its rotation gates), the constant adder $\Phi$ADD has a depth of 1, the controlled C$\Phi$ADD and the adder $\Phi$ADD have a depth of $n$  and the $\Phi$MAC as analyzed in the previous section has a depth of $8n$ . In this depth analysis we have also taken into account that many of the blocks can be executed in parallel. Fault tolerance and implementation aspects are discussed in a separate paragraph in Section \ref{sec:Analysis}.

\section{Generic Modular Multipler/Accumulator and Modular Multiplier}\label{sec:FMOD1}
\noindent

In this Section we build on the previous blocks two quantum circuits, the generic QFT based controlled modular multipler/accumulator and then we proceed to a generic quantum controlled modular multipler which can be used as the basic building block for the modular exponentiation circuit as shown in Figure \ref{fig:HighLevelShor1qubit}. 

\subsection{Generic Controlled QFT Modular Multiplier/Accumulator - {\textbf{\textit \textPhi}}MAC\_MOD1}
\noindent
Controlled modular multipliers/accumulators can be built using the $\Phi$MAC and GM$\Phi$DIV1 units as their basic blocks. Such blocks can realize Eq. \ref{eq:FMACMOD1}, and then can be used as described in Section 5.2 to realize a controlled modular multiplier as that of Eq. \ref{eq:CUblock}. When the multiplicand input of the $\Phi$MAC unit is $n$ qubits wide its accumulator is $2n$ qubits wide. These $2n$ qubits must be fed as input to the dividend input of the GM$\Phi$DIV1 to compute the modular result. Therefore, the size of the GM$\Phi$DIV1 unit must such that it can receive a dividend of $2n$ qubits, which means that the dividend input must be $4n$ qubits wide and its upper half $2n$ qubits should be zero. Therefore the required ancillae number for the GM$\Phi$DIV1 grows from $5n+1$ to $10n+1$. Note that a simple interconnection of the two units in succession is not adequate to give a result that follows Eq. \ref{eq:FMACMOD1} because the GM$\Phi$DIV1 unit gives at its outputs both the remainder which is the useful part of the computation and the quotient which contains useless qubits that must be reset to keep the reversibility of the circuit and can be reused at a later time. 

\begin{equation}
\mathit{\Phi} MAC\_MOD1_{a^{2^{i}},N}  (|c\rangle |y\rangle |0\rangle) = |c\rangle |y\rangle |( a^{2^{i}})^{c} y \bmod N \rangle 
\label{eq:FMACMOD1}
\end{equation}

\begin{figure} [htb]
\centerline{\epsfig{file=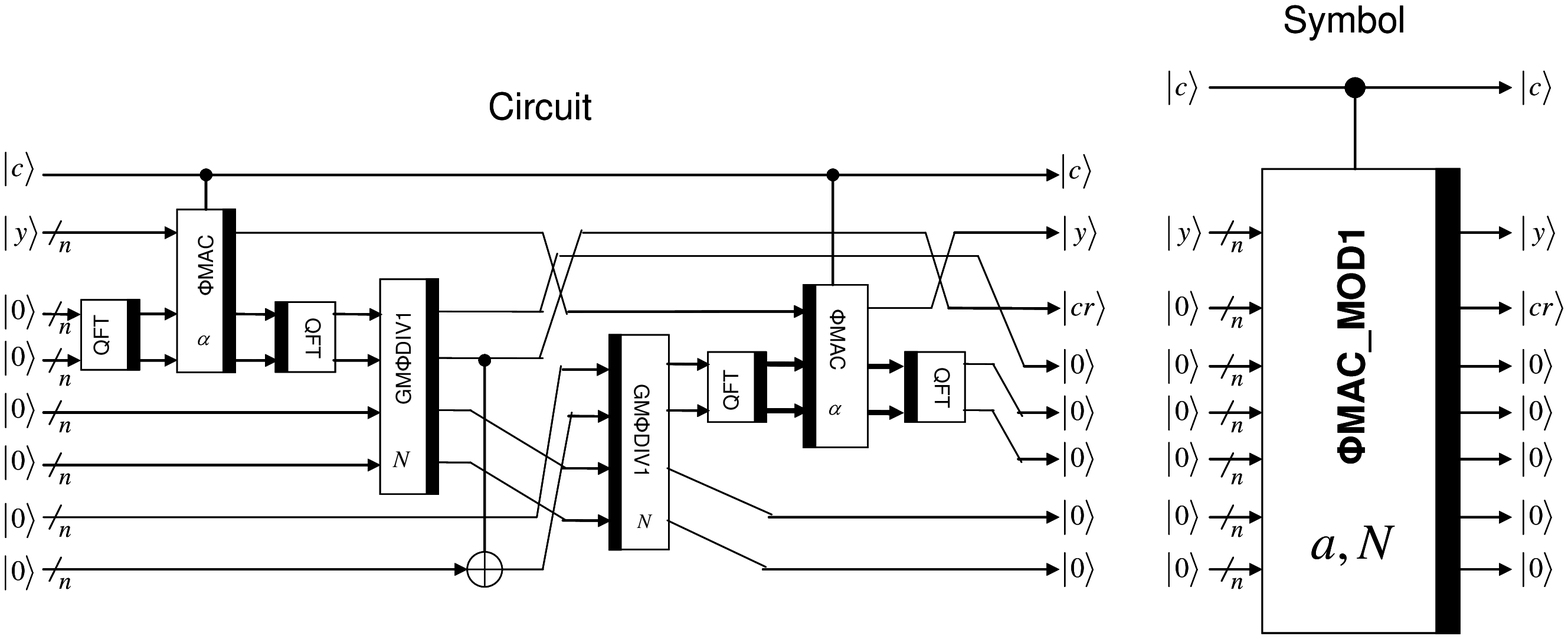, width=13cm}} 
\vspace*{13pt}
\fcaption{\label{fig:FMACMOD1} The full diagram of the generic controlled modular multiplier/accumulator $\Phi$MAC\_MOD1 and its symbol. A total of $7n+1$ qubits are shown here, but there are $10n+1$ more ancilla qubits not shown in the GM$\Phi$DIV1 symbol.  }
\end{figure}

The proposed architecture of the generic controlled modular multiplier/accumulator by constant, named $\Phi$MAC\_MOD1, is shown in Figure \ref{fig:FMACMOD1}. This circuit diagram shows $7n + 1$ qubits, but there are $10n + 1$ more “hidden” qubits in the GM$\Phi$DIV1 symbol (the ancillae of the first divider can be reused in the second one, as the second one cannot be operated in parallel to the first) which we don't show for the clarity of Figure \ref{fig:FMACMOD1}. The input lines with a slash symbol in the figure correspond to “buses” of $n$ qubits. It is straightforward for the reader to understand the equivalence between the symbols with individual qubits presented in previous sections with the ones in Figure \ref{fig:FMACMOD1} having buses as inputs and outputs. Apart from the two basic blocks and the QFT units there is a group of CNOT gates in this diagram. These units operate on a “qubitwise” basis, i.e. the first qubit of the controlling bus controls the CNOT of the first qubit of the target bus, etc. An analysis of this circuit follows for the two cases of the controlling qubit $|c \rangle =|1\rangle $ and $|c \rangle =|0\rangle $. As in the previous Section, we will not care of the various QFT blocks present in the register buses and we give the integers values as not being QFT transformed.

For the case of $|c \rangle = |1 \rangle $   both $\Phi$MAC units are enabled. Initially all the $n$ ancilla qubit buses are in state $|0 \rangle $   while the value $y$, the number to be multiplied by the constant $a$, is fed to the multiplicand input of the first $\Phi$MAC($a$) unit. After the operation of this $\Phi$MAC($a$) unit the multiplicand qubits are still $|y \rangle $  while the $2n$ qubits of its accumulator go to state  $|0+ay \rangle =|(ay)_{U}\rangle |(ay)_{L}\rangle$, where subscripts $U$ and $L$ denote upper and lower qubits, respectively. These $2n$ qubits feed the dividend input of the GM$\Phi$DIV1($N$) giving as outputs the remainder  $|(ay \bmod N)_{U}\rangle |(ay \bmod N)_{L}\rangle = |0 \rangle |(ay \bmod N)_{L}\rangle $ and the quotient $ |q_{U}\rangle |q_{L}\rangle $, where $N$ is again the number to be factored. It is assured that the upper $2n-n=n$ qubits of the remainder are zero because $N$ is $n$ bits wide. The remainder is copied with the aid of the group of the CNOT gates to an $n$ qubits bus and is fed to an inverse GM$\Phi$DIV1($N$) unit whose other input is the quotient $| \lfloor(ay)/N \rfloor \rangle$ computed by the first GM$\Phi$DIV1($N$)  unit. Such an inverse divider can be easily designed by reversing the signal flow of the normal divider and setting the angle in every rotation gate of the inverted GM$\Phi$DIV1 to the opposite of the original angle. By feeding the quotient and the remainder in an inverse GM$\Phi$DIV1 we take as output the input that would give this remainder and quotient, that is $|ay \rangle $ at the top $2n$ qubits and $|0 \rangle$ at the bottom $2n$ qubits (an inverse divider effectively becomes a multiplier). The second $\Phi$MAC($a$) unit, which is inverted, takes  as multiplicand input the state $|y \rangle$ and as accumulator input the output of the previous inverted GM$\Phi$DIV1($N$), $|ay \rangle$. Similarly, the outputs of this inverted $\Phi$MAC($a$) are the inputs that would lead a normal $\Phi$MAC unit to give as outputs the inputs being fed to the inverted, that is $|y \rangle $ at the top $n$ qubits and  $|0 \rangle $   at the lower $2n$ qubits. This way we achieved to clear the useless, in our application, quotient. At this point, we have the desired remainder $|(ay \bmod N) \rangle$ available along with the initial input $|y \rangle $.

The case of setting the control qubit $|c \rangle =|0\rangle $   is simpler than the previous one. Both the $\Phi$MAC units are disabled and they simply pass their inputs unmodified to their outputs. Therefore, input $|y \rangle $   traverses the circuit through the $\Phi$MAC units while the ancilla buses remain in the zero state.

\subsection{Generic Controlled QFT Modular Multiplier - {\textbf{\textit \textPhi}}MUL\_MOD1}
\noindent

\begin{figure} [t]
\centerline{\epsfig{file=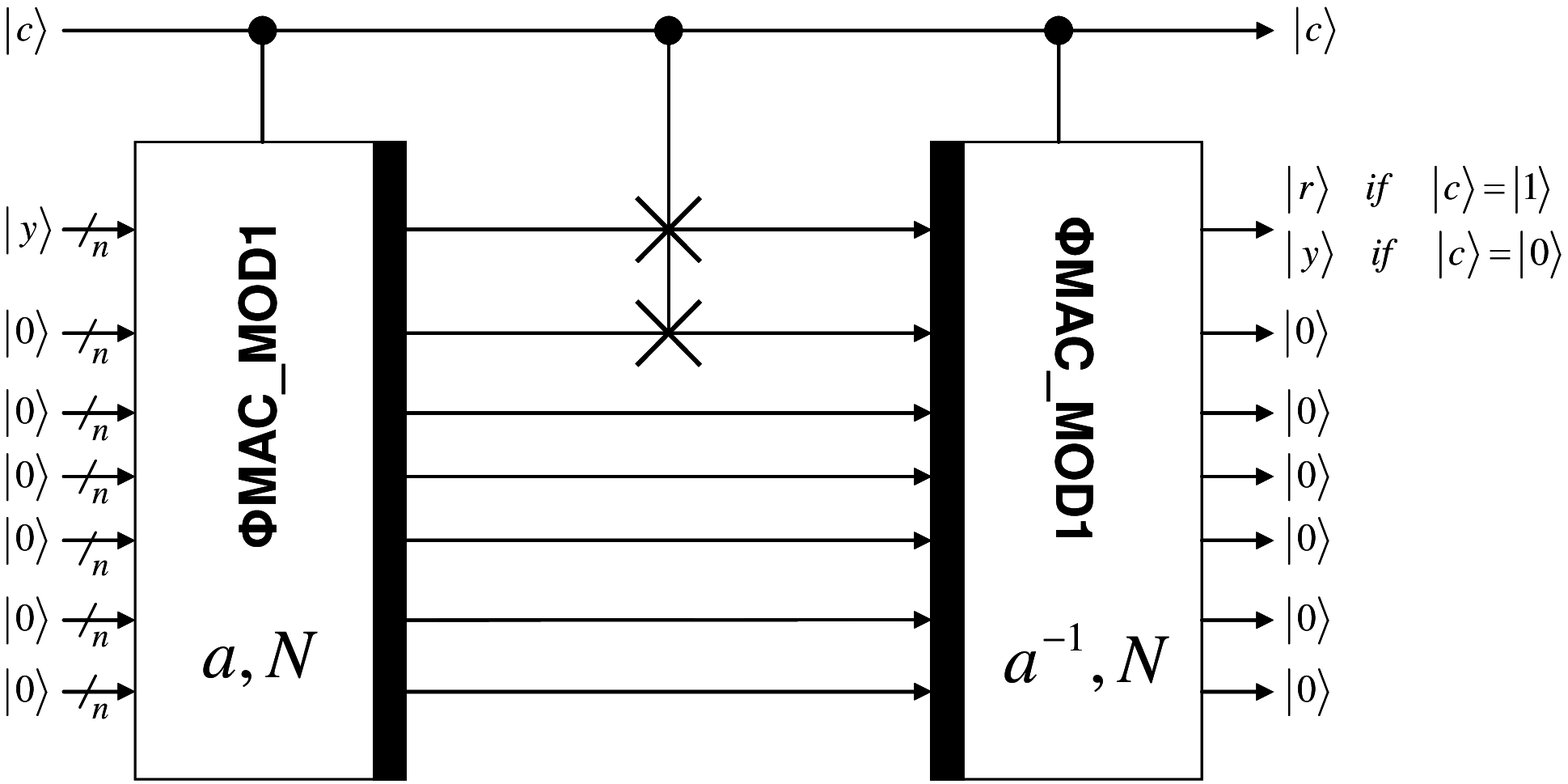, width=8cm}} 
\vspace*{13pt}
\fcaption{\label{fig:FMULMOD1} Generic modular multiplier $\Phi$MUL\_MOD1. The circuit requires the $7n+1$ qubits shown in the diagram, plus $10n+1$ ancilla qubits hidden in the divider units.  }
\end{figure}

The last step in the construction of the modular multiplier required by Shor's algorithm is to clear the state $|y \rangle$   at the output of the modular multiplier/accumulator to the zero state so that we can successively connect several modular multiplier units as shown in Figure \ref{fig:HighLevelShor1qubit}. Figure \ref{fig:FMULMOD1} shows a method for clearing the undesired $|y \rangle$ \cite{Beauregard}. Two $\Phi$MAC\_MOD1 units are used in this diagram, with a block of controlled SWAP gates (Fredkin gates). The second $\Phi$MAC\_MOD1 unit is a reverse unit with multiplication parameter $a^{-1}$, where the inverse $a^{-1}$  is defined with respect to the operation of multiplication modulo $N$, that is it must hold $a \cdot a^{-1} (\bmod N) = 1$. Such an inverse always exists, because the randomly picked number $a$ is selected based on the restriction that it must be co-prime with $N$. The analysis of this circuit for the case of $|c \rangle = |0\rangle $ is very simple as both the $\Phi$MAC\_MOD1 units and the CSWAP gates are disabled. Thus, input state $|y \rangle$   remains unmodified and passes to the output, while all the ancilla qubits remain in the zero state. In the case of $|c \rangle = |1\rangle $, the first $\Phi$MAC\_MOD1($a,N$) unit gives as result the input $|y \rangle$ and the remainder $|r \rangle = |ay \bmod N\rangle$. This remainder is then fed, through the CSWAP gates, to the multiplicand input of the second $\Phi$MAC\_MOD1($a^{-1},N$) unit, while the accumulator input of this second unit is $|y \rangle$. This way the multiplicand output of $\Phi$MAC\_MOD1($a^{-1},N$) becomes the remainder  $|r \rangle$ while the accumulator becomes $| y-a^{-1}( ay \bmod N  ) \bmod N \rangle = | y-y \rangle = | 0 \rangle $.

Combining the two case of  $|c \rangle = |0\rangle $ and $|c \rangle = |1\rangle $   we have the general case transformation:

\begin{equation}
\mathit{\Phi} MUL\_MOD1_{a,N}  (|c\rangle |y\rangle |0\rangle |0\rangle |0\rangle |0\rangle |0\rangle |0\rangle) = |c\rangle |a^{c}y (\bmod N) \rangle |0\rangle |0\rangle |0\rangle |0\rangle |0\rangle |0\rangle 
\label{eq:FMULMOD1}
\end{equation}

which has exactly the same form as Eq. \ref{eq:CUblock}, if the ancilla qubits are not taken into account. The final result is that we can combine many $\Phi$MUL\_MOD1 units of Figure \ref{fig:FMULMOD1} as shown in Figure \ref{fig:HighLevelShor1qubit} to build a quantum modular exponentiation circuit.

\section{Optimized Modular Multipler/Accumulator and Modular Multiplier}\label{sec:FMOD2}
\noindent
Exploitation of the specific application where the modular multiplier is to be used can be advantageous in terms of both depth and qubits requirements. The following paragraphs present  optimized versions of the modular multiplier/accumulator and modular multiplier.

The second version of the modular multiplier/accumulator, which we denote as $\Phi$MAC\_\-MOD2, exploits the specific application to be used to, namely Shor's factorization algorithm. As shown in Figure \ref{fig:HighLevelShor1qubit}, each modulo $N$ multiplier unit takes as input the output of its previous unit which is again a modulo $N$ multiplier block (or the integer 1 for the first unit) and thus this input is always less than $N$.  This input is to be multiplied by an integer   which is again always less than $N$. Therefore, the product of these two integers, being less than $N^{2}$, has to be divided by $N$ to calculate the remainder. The quotient of this division is of course again less than $N$. Taking as $n=\lceil log_{2}N \rceil$   the number of qubits for the division circuit we can see that for this specific case the quotient is less than $2^{n}$. In other words, the restriction imposed for the operation of the Granlund-Montgomery division algorithm holds. For this reason we can use the division circuit of Figures \ref{fig:GMFDIVa}-\ref{fig:GMFDIVb} with a size of only $n$ bits, that is a dividend of $2n$ qubits (the upper half is not necessary to be zero) and quotient and remainder sizes of $n$ qubits, instead of using the double sized generic divider of the previous section. 

\begin{figure} [t]
\centerline{\epsfig{file=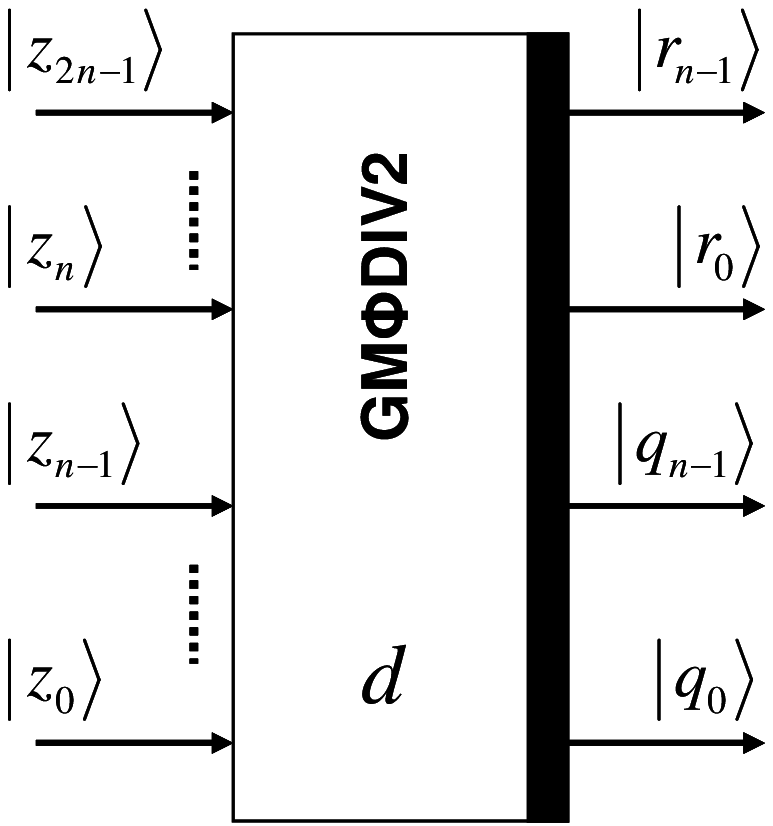, width=4cm}} 
\vspace*{13pt}
\fcaption{\label{fig:GMFDIV2} Symbol of GM$\Phi$DIV2 that receives a dividend of $2n$ qubits, subject to the constraint that the quotient is less than $2^{n}$. The underlying circuit uses $7n+1$ qubits.  }
\end{figure}

\begin{figure} [!b]
\centerline{\epsfig{file=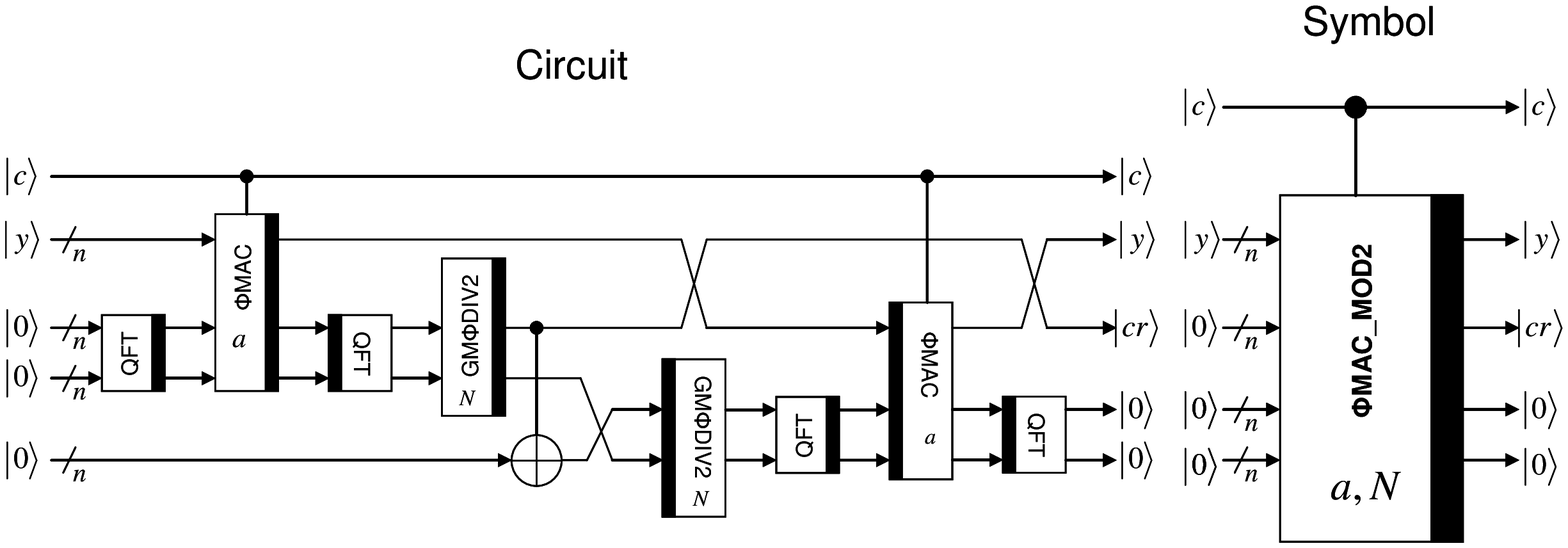, width=13cm}} 
\vspace*{13pt}
\fcaption{\label{fig:FMACMOD2} The optimized controlled modular multiplier/accumulator $\Phi$MAC\_MOD2 and its symbol. A total of $4n+1$ qubits are shown in this figure, but there are $5n+1$ more ancilla qubits not shown in the GM$\Phi$DIV1 symbol.     }
\end{figure}

\FloatBarrier

We introduce a new symbol for the same divider in Figure \ref{fig:GMFDIV2}, merely reflecting the fact that all of the $2n$ qubits are reserved for the dividend, in contrast to the symbol of Figure \ref{fig:GMFDIV1} where half of them where set to zero in order to comply with the restriction of the quotient being less than $2^{n}$. All the other internal aspects of the GM$\Phi$DIV2 are the same as of GM$\Phi$DIV1. This second version of the divider is to be used exclusively in a Shor's quantum algorithm architecture where the quotient is expected to be always less than $2^{n}$, while the first version of the divider can be used whenever a general quantum divider by constant is needed.

The proposed architecture of the second version of controlled modular multiplier / \-accumulator by constant, named $\Phi$MAC\_MOD2, is shown in Figure \ref{fig:FMACMOD2}. The circuit diagram shows a total of $4n + 1$ qubits but there are $5n + 1$ more “hidden” qubits in the GM$\Phi$DIV2 symbols which are not shown for the shake of clarity of Figure \ref{fig:FMACMOD2}. Again, the input lines with a slash symbol in the figure correspond to buses, each bus consisting  of  $n$ qubits. Using similar arguments as in the previous section we give a brief analysis of this circuit.

For the case of $|c \rangle = |1\rangle $ both $\Phi$MAC units are enabled. The accumulator register output of the first $\Phi$MAC($a$) unit is in state $ |ay\rangle _{U} |ay\rangle  _{L}$,  while its  multiplicand register still holds the multiplicand $|y \rangle$. The product $ |ay\rangle _{U} |ay\rangle  _{L} $ is then fed to the first GM$\Phi$DIV2($N$) to produce the remainder (upper bus of GM$\Phi$DIV2($N$)) and the quotient (lower bus of GM$\Phi$DIV2($N$)). The remainder is copied to the lower ancilla bus of the circuit and both the quotient and the copied remainder are fed to an inverted GM$\Phi$DIV2($N$) unit, giving again $ |ay\rangle _{U} |ay\rangle  _{L}$ at its output. The second (inverted) $\Phi$MAC($a$) unit has then at its multiplicand input the $|y\rangle$ and at its accumulator input $ |ay\rangle _{U} |ay\rangle  _{L}$, setting its accumulator output to state $|0 \rangle$. Now, as in the case of $\Phi$MAC\_MOD1, we have the desired remainder $|(ay \bmod N) \rangle$ available along with the initial input $|y \rangle$. Similarly, the case of  $|c \rangle = |0\rangle $ leads the top qubits bus to have the initial input $| y \rangle$, while all the other buses are set to the zero state.

\begin{figure} [!t]
\centerline{\epsfig{file=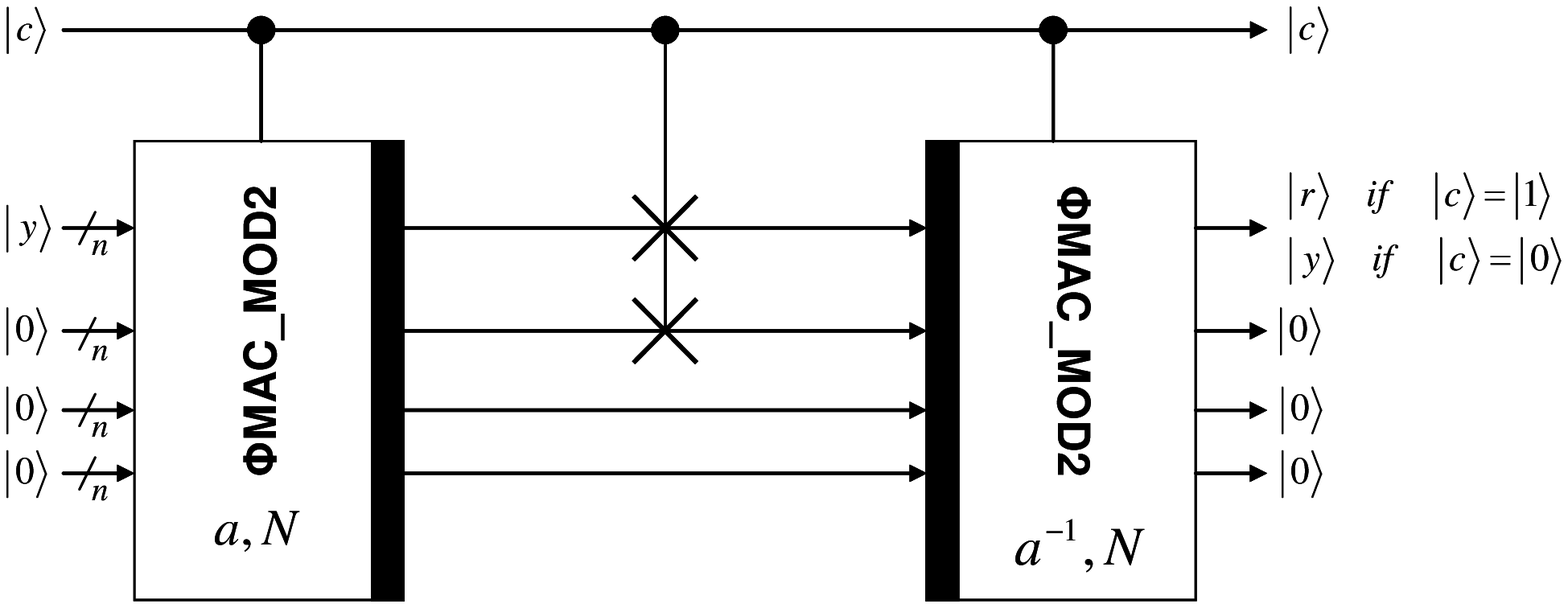, width=8cm}} 
\vspace*{13pt}
\fcaption{\label{fig:FMULMOD2} Optimized modular multiplier $\Phi$MUL\_MOD2. The circuit requires the $4n+1$ qubits shown in the diagram, plus $5n+1$ ancilla qubits hidden in the divider units.     }
\end{figure}

A design of a modular multiplier based on the optimized multiplier/accumulator $\Phi$MAC\_\-MOD2 is shown in Figure \ref{fig:FMULMOD2}. This design is very similar to the generic case of Section 5.2 using the generic Φ$\Phi$MAC\_MOD1 unit and for this reason we don't analyze the circuit.

\FloatBarrier

\section{Complexity Analysis}\label{sec:Analysis}
\noindent
In this section we analyze the depth (speed), width (the required number of qubits) and the quantum cost (total number of equivalent one or two qubit gates) of a quantum modular exponentiation circuit, when implemented using either the $\Phi$MUL\_MOD1 unit, or the optimized $\Phi$MUL\_MOD2 unit. We also compare the two proposed designs with other designs found in the literature, in terms of depth and width.
We begin the depth analysis for the case of the $\Phi$MUL\_MOD1 controlled modular multiplier as top level unit for the modular exponentiator. The $\Phi$MUL\_MOD1 block consists of the two $\Phi$MAC\_MOD1 units and controlled SWAP gates (CSWAP), while each $\Phi$MAC\_MOD1 unit consists of the units QFT, $\Phi$MAC, GM$\Phi$DIV1 whose depth is already analyzed and CNOT gates.  As it is shown in Figure \ref{fig:FMACMOD1} none of these units can operate in parallel, because each one gets its inputs from the previous one, so the depth of this modular multiplier/accumulator is merely the sum of the depth of each unit. We note here that the depth of the CNOT unit used in $\Phi$ΦMAC\_MOD1 is 1 since the gates of this unit can operate in parallel and the depth of the CSWAP unit used in $\Phi$MUL\_MOD1 is about 5n (a CSWAP gate consists of two CNOT gates and a Toffoli gate). The QFT size is for $2n$ qubits and also the GMΦDIV1 size is for $2n$ bits dividend even if the divisor $N$ is $n$ bits wide, therefore the depths of the QFT and GM$\Phi$DIV1 units used in the $\Phi$MUL\_MOD1 unit are  $2\cdot 2n - 1 = 4n - 1$ and $244 \cdot 2n - 8 = 488 n - 8$, respectively. In Table 3 we summarize the depth of each unit used in the modular multiplier unit (we have take into account that each multiplier unit uses two multiplier/accumulator units), the number of units of each type that are used, the depth contribution of each type of unit and finally we calculate the total depth for the controlled modular multiplier $\Phi$MUL\_MOD1 which is $2021n-38$. Since the complete modular exponentiator circuit consists of $2n$ $\Phi$MUL\_MOD1 blocks successively connected in series, its depth is about $2n \cdot 2021n=4042n^{2}$.
As for the number of qubits required for the complete modular exponentiator circuit ($\Phi$EXP1), we see that in the case of the $\Phi$MUL\_MOD1 unit, it requires the same number of qubits as the controlled modular multiplier, that is $7n + 1$ qubits plus the hidden ancilla bits of the GM$\Phi$DIV1 not shown in Figure \ref{fig:FMACMOD1}. The ancilla qubits for a GM$\Phi$DIV1 unit of size $2n$ is $10n + 1$ qubits, so the total qubits required for the circuit is $17n + 2$.
The quantum cost for the $\Phi$MUL\_MOD1 modular multiplier calculated from Table \ref{table:MOD1cost} is about $2900n^{2}$, and thus the cost of the first modular exponentiation circuit is about $5800n^{3}$.

\vspace*{4pt}   
\begin{table}[t]
\tcaption{Units used in the $\Phi$MUL\_MOD1 design, depth of each unit, number of gates in each unit,  number of units used for each type, gates contribution and depth contribution of each type of unit to the total quantum cost and depth.}
\centerline{\footnotesize\smalllineskip
\begin{tabular}{l c c c c c}\\
\hline
Unit & Depth per unit & Cost per unit     & \# of units & Cost & Depth \\\hline
QFT($2n$) & $4n-1$ & $10n(n+1)$ & 8 & $80n^{2}+80n$ & $32n-8$ \\
$\Phi$MAC($n$) &	$8n$ &	$4n(n+1)$	 & 4	& $16n2+16n$	& $32n$ \\
GM$\Phi$DIV1($2n$)	& $488n-8$ & 	$700n^{2}+298n $ & 	4	& $2800n^{2}+1196n$ &	$1952n-32$ \\
CNOT($n$)	& 1 &	$n$ &	2	& $2n$ &	2 \\
CSWAP($n$) &	$5n$	& $5n$	& 1	& $5n$	& $5n$ \\\hline
\multicolumn{4}{l}{Total} & $2896n^{2}+1299n$ &	$2021n-38$ \\
\hline
\end{tabular}}
\label{table:MOD1cost}
\end{table}

A similar depth analysis for the second, optimized proposed design of the complete modular exponentiator $\Phi$EXP2 that utilizes the $\Phi$MUL\_MOD2 is shown in Table \ref{table:MOD2cost}. When the optimized modular multiplier is used we get an improved depth of about $2n \cdot 1045n = 2090n^{2}$, that is half the depth of the case of using the generic $\Phi$MUL\_MOD1 multiplier. However, the most important gain in using the $\Phi$MUL\_MOD2 is that the required qubits are only the $4n + 1$ qubits shown in Figure \ref{fig:FMACMOD2} plus the $5n + 1$ qubits hidden inside the symbol of GM$\Phi$DIV2, that is a total of $9n + 2$, a significant improvement over the $17n + 2$ qubits if using the  $\Phi$MUL\_MOD1 modular multiplier requirement. 

Similarly, the quantum cost for the $\Phi$MUL\_MOD2 modular multiplier is about $800n^{2}$, hence the quantum cost for the modular exponentiator circuit is about $1600n^{3}$.

We compare the two proposed designs to various quantum modular exponentiation circuit designs found in the literature. Table \ref{table:comparison} shows the abbreviated names of the basic building blocks for each exponentiation circuit as given in subsection 2.2, a description of its main building block, the number of qubits required for the full modular exponentiation circuit and an estimation of its depth. The two circuits proposed in this paper are referred in Table \ref{table:comparison} as $\Phi$EXP1 and $\Phi$EXP2. The design $\Phi$EXP1 is the modular exponentiation circuit built from the generic components GM$\Phi$DIV1 and $\Phi$MAC\_MOD1 and has inferior performance to the $\Phi$EXP2 because it doesn't exploit the property mentioned in Section \ref{sec:FMOD2}. It appears to the comparison table just for reference.

\vspace*{4pt}   
\begin{table}[t]
\tcaption{Units used in the $\Phi$MUL\_MOD2 design, depth of each unit, number of gates in each unit, number of units used for each type, gates contribution and depth contribution of each type of unit to the total quantum cost depth.}
\centerline{\footnotesize\smalllineskip
\begin{tabular}{l c c c c c}\\
\hline
Unit & Depth per unit & Cost per unit     & \# of units & Cost & Depth \\\hline
QFT($2n$) & $4n-1$ & $10n(n+1)$ & 8 & $80n^{2}+80n$ & $32n-8$ \\
$\Phi$MAC($n$) &	$8n$ &	$4n(n+1)$	 & 4	& $16n2+16n$	& $32n$ \\
GM$\Phi$DIV2($n$)	& $244n-8$ & 	$175n^{2}+149n $ & 	4	& $700n^{2}+596n$ &	$976n-32$ \\
CNOT($n$)	& 1 &	$n$ &	2	& $2n$ &	2 \\
CSWAP($n$) &	$5n$	& $5n$	& 1	& $5n$	& $5n$ \\\hline
\multicolumn{4}{l}{Total} & $796n^{2}+629n$ &	$1045n-38$ \\
\hline
\end{tabular}}
\label{table:MOD2cost}
\end{table}

Regarding the circuit depth, it is referred to the depth of one-qubit gates or two-qubits gates. Whenever three-qubit gates are encountered (e.g. Toffoli gates) their depth is rough approximated  to an equivalent depth of one-qubit or two-qubits gates by assuming each three-qubit gate can be replaced by five one-qubit or two-qubits gates.  Not all designs of the literature provide a full circuit; thus some estimations are rough and are based on \cite{VanMeter1,Choi1} and our assumptions. For example to calculate the depth of a full exponentiation circuit based on an particular adder if a full circuit for the exponentiation operation is not given, we made the assumption that a modular adder for the exponentiation circuit needs five normal adders to be built. At this point we have to warn that some depths as referred by some authors are not to be taken as is, as these authors tend to make their calculation not by counting one-qubit, two qubit gates and converting the depth of the Toffoli gates to an equivalent depth, but they rather group together various gates and count computation steps for each group. Also, in some of the previous works the depth is calculated as two qubit interactions, e.g. an arbitrary sequence of gates on two adjacent qubits is measured as having depth 1. This assumption, depending on the specific physical implementations, may not hold. For the above reasons (not full or detailed circuit given, adjacent qubit gates depth assumption) in some entries of Table \ref{table:comparison} we give measures in $O(\cdot)$ notation  instead of expressions with leading order constants. 

\vspace*{4pt}   
\begin{table}[t]
\tcaption{Comparison of various modular exponentiation circuits in terms of qubits requirement and asymptotic depth (speed) behaviour, where Toffoli gates are assumed to contribute five times the depth of two or one qubit gates.}
\centerline{\footnotesize\smalllineskip
\begin{tabular}{l l l l  l}\\
\hline
Basic Block & Type of adder used & \#qubits    &  Depth \\\hline
VBE \cite{Vedral}  & Ripple Carry & 	$7n+1$&	 $\sim 500n^{3}$ \\
BCDP \cite{Beckman} & Ripple Carry &	$5n+3$	&  $\sim 280n^{3}$ \\
CDKM \cite{Cuccaro} & Ripple Carry &	 $\sim 4n$ &	$\sim 200n^{3} $ \\
DKRS \cite{Draper2} & Carry look-ahead &	$\sim 6n$ &	$\sim 400n^{2}log_{2}n$ \\
TK \cite{Takahashi}  & Ripple Carry 	   & $\sim 3n$	& $\sim 500n3$ \\ 
VI-algorithmD \cite{VanMeter1}  & Conditional Sum	& $2n^{2}$	& $\sim 45n(log_{2}n)^{2}$ \\
VI-algorithmE \cite{VanMeter1} & DKRS 	& $2n^{2}$ & 	$\sim 55n(log_{2}n)^{2}$ \\
VI-algorithmF \cite{VanMeter1} & CDKM 1D-NTC	& 	$2n^{2}$ &	$\sim 100n^{2}log_{2}n$ \\
Beauregard \cite{Beauregard} & QFT adder	& $2n+1$ &	$\sim 100n^{3}$ \\ 
Gosset \cite{Gossett} & Carry Save &	~$8n^{2}$	& $O(nlog_{2}n)$ \\ 
Zalka 1 \cite{Zalka} & Carry Select Approximate	 & $5n$	& $\sim 3000n^{2}$ \\
Zalka 2 \cite{Zalka} & FFT Multiplier	& $24n \ldots 96n$ &	$\sim 2^{19}n^{1.2}$ \\
FDH \cite{Fowler1} & QFT adder 1D-NTC	& $\sim 2n$	& $O(n^{3})$ \\ 
Kutin 1 \cite{Kutin} & QFT adder/Approximate 1D-NTC &	$\sim 3n$	& $O(n^{2})$ \\
Kutin 2 \cite{Kutin} & CDKM/Approximate 1D-NTC	& $\sim 3n$	& $O(n^{2}log_{2}n)$ \\ 
CV  \cite{Choi1} & Carry look-ahead 2D-NTC	& $\sim 4n$ &	$\sim 750n^{2}\sqrt{n}$ \\ 
PS  \cite{Pham} & Carry Save 2D-NTC	& $O(n^{4})$	& $O((log_{2}n)^{2})$ \\ 
$\Phi$EXP1 (proposed)	& QFT Adder/MAC/Div & 	$17n+2$ &	$\sim 4000n^{2}$ \\
$\Phi$EXP2 (proposed)	& QFT Adder/MAC/Div	& $9n+2$ &	$\sim 2000n^{2}$ \\
\hline
\end{tabular}}
\label{table:comparison}
\end{table}

We can see in Table \ref{table:comparison} that the $\Phi$EXP2 circuit outperforms in terms of speed in most cases all the circuits that are based on ripple carry adders, carry look-ahead adders and  outperforms Beauregard’s and Fowler-Devitt-Hollenberg QFT based circuits,  while at the same time requires qubits of the same size order. 

Algorithms D, E and F of Van Meter and Itoh (VI) try to improve the depth by applying various techniques such as better depth modulo calculation, indirection \cite{VanMeter1} but the main improvement is done by operating in parallel many modular multipliers at the cost of a respective increase of the qubits. As many tuning parameter are used in the VI algorithms, the expressions giving the qubits number and the depth are complicated. In Table \ref{table:comparison} we show the asymptotic depth when the largest number of qubits ($2n^{2}$) can effectively be used by these algorithms \cite{VanMeter2} that is when we take advantage of the highest offered concurrency. In this case algorithms D and E can achieve a depth of $O(n log_{2}n)$ which is asymptotically better than the proposed $\Phi$EXP2 design but this improvement will require $O(n^{2}$)  space. Algorithm F has for the same number of qubits ($2n^{2}$) a depth of $100n^{2}log_{2}n$ which is asymptotically worst than the proposed design. If we relax the space requirements of the three VI algorithms D, E and F to be linear $O(n)$, then algorithms D and E offer an asymptotical depth $O((nlog_{2}n)(n/s+log_{2}s))$ where $s$ is the number of concurrent multipliers used and algorithm F offers a depth of $O((nlog_{2}n)(n/s+log_{2}s))$. These depths are worst in order than the asymptotical depth of the $\Phi$EXP2 architecture if $s$ is set constant. 

Gosset's carry save adder circuit has smaller depth than $\Phi$EXP2 circuit but with a large penalty in space because the number of qubits bits it requires depends quadraticaly on the size of the number to be factored.

The same applies for the Pham and Svore (PS) two-dimensional architecture which has a $O(log_{2}n)^{2}$  depth but requires a tremendous $O(n^{4})$  space. The second two-dimensional architecture (CV) which requires space of about $4n$ has also asymptotically worst depth of $O(n^{2}\sqrt{n})$.

Zalka's FFT multiplier circuit performs better than $\Phi$EXP2 but only for numbers to be factored with more than 10kbits in size due to its big constant in its depth. Also it uses much more qubits than $\Phi$EXP2. Zalka's first circuit (using carry select adder) is comparable to $\Phi$EXP2 in terms of both depth and qubits required, the $\Phi$EXP2 being faster but using twice the qubits number of Zalka's circuit. At this point we have to mention that Zalka's first circuit makes only approximate calculations of the modular exponentiation function. 

The other architecture which has comparable asymptotical depth to $\Phi$EXP2 is Kutin's one-dimensional architecture based on QFT addition, being an approximate calculation circuit like Zalka's first circuit. The second one of Kutin's circuit is slightly worse in terms of depth   and makes the same approximation as its first.

In conclusion, the $\Phi$EXP2 circuit has the lowest asymptotical depth among the circuits that require a linear number of qubits smaller than $10n$ and are based on exact (as opposed to approximate) calculations.

The $\Phi$EXP2 circuit natively uses almost exclusively two-qubit gates (the only exception are the CSWAP gates which use Toffoli gates). This is an advantage over most of the architectures of Table \ref{table:comparison} (apart from Beauregard's circuit which can be also transformed to use almost exclusively two qubit gates) because physical implementations of quantum gates of three qubits is difficult \cite{DiVincenzo}. Even recent proposals of Toffoli gate implementation in various technologies \cite{Lanyon,Ralph,Monz,Fedorov}  essentially resolve this problem by decomposition into two qubit gates.

On the other hand, concerning the implementation, the proposed architecture has two major weaknesses. The first comes from the fact that it uses extensively controlled rotation gates and there are known problems concerning the fault tolerance of such gates \cite{VanMeter1}. In general, the rotation gates are not included in the family of gates that can be realized in a fault tolerant manner using known quantum error correcting codes, e.g. Steane codes. But even if they are not inherently fault tolerant they can become such, if a decomposition into fault tolerant gates is applied to them exploiting the Solovay-Kitaev theorem \cite{Kitaev}. 

The problem of decomposing an arbitrary quantum single qubit gate into a pre-determined set of fault tolerant gates is important because for such a decomposition a cost have to be paid related to the number of the sequence of gates to be used to realize the decomposition and consequently it is related to a depth increment of the total circuit. The Solovay-Kitaev theorem states that given a small constant $\epsilon$, an arbitrary gate $U$ can be approximated by a finite sequence of gates equivalent to a gate $S$ up to an approximation error $\epsilon$ (that is $d(U,S)<\epsilon$, where $d$ is a distance function), the length of this sequence being $O(log^{c}(1/\epsilon))$. The constant $c$ is somewhere between 1 and 2. As the smallest rotation angle in the rotation gates of the proposed architecture is of the order $\pi/2^{n}$, the required approximation error is of the same order (otherwise we would use the identity gate instead \cite{Fowler2}) and the depth cost is of the order $n^{c}$.

Much research has been done in the area of this problem. Some of this work is oriented to improve the constant factor $c$ that is to improve the cost and depth of the decomposed circuit. In \cite{Fowler2,Fowler3} a decomposition of a controlled rotation gate into two single qubit rotation gates and a CNOT gate is applied, and then the single qubit rotation gate is approximated, up to an error constant $\epsilon$, by a sequence of $H$ and $T$ ($\pi/8$) gates. It is shown that this approximation has a length linear in $O(log(1/\epsilon))$, that is the constant $c$ is equal to 1, but the major drawback of this method is the synthesis time required, which is exponential in $1/\epsilon$.

Very recently, a very intense research activity has been observed in the area of efficiently synthesizing arbitrary unitaries using fault tolerant primitive gates. Some work is oriented in trading-off circuit complexity with synthesizing time complexity, other work is diverted between building arbitrary gates with and without ancillae, and some other work is directed to using not standard primitive gate set. A non exhaustive list of such work includes \cite{Trung,CodyJones1,Selinger,Kliuchnikov,Bocharov}. In summary, these results show that an arbitrary gate can be approximated with linear length in its accuracy $1/\epsilon$ and with linear synthesis time in $1/\epsilon$. The constant in the complexity of the approximation circuit is about 10 for the most promising proposals. We think, as these results suggest, that the proposed architecture for modular exponentiation could be a candidate architecture in the future if someone gives priority to depth.

The second weakness of the proposed architecture is that it does not account for the possible constraints on the communication distance between the qubits that may imposed by the underlying physical implementation. For example in \cite{Choi2} is shown that the physical mapping of any quantum adder of $\Omega(log_{2}n)$ depth to a $k$-dimensional nearest neighbors architecture limits its theoretical depth to $\Omega(\sqrt[k]{n})$. Such depth limitations are due to the additional SWAP gates needed to convert from the abstract concurrent architecture to the nearest neighbor architecture.
 
Other models of quantum computation more suitable for long distance communication between gates are the measurement based quantum computation (MBQC) \cite{Raussendorf} where the interaction between distant qubits can be accomplished in constant time. In \cite{Trisetyarso}  the DKRS carry look-ahead adder is redesigned in the MBQC context and it is shown that its logarithmic depth can be maintained but with a substantial overhead in space requirements.

Surface code quantum computing is another promising scalable architecture which also permits long distance interactions \cite{Fowler4,VanMeter3,CodyJones2}. In general, the estimation of area and speed of a quantum algorithm is a cumbersome task when considering the real physical implementation of the algorithm, including the error-correcting codes which achieve the required fault-tolerance. In this view, the comparison between the various designs exposed in Table \ref{table:comparison} is only indicative for their performance.
 
It is important to note that the $\Phi$EXP2 circuit can be made about three times faster if the \emph{approximate} QFT method of performing the additions \cite{Draper1,Barenco2,Coppersmith} is utilized. Unfortunately, the approximate QFT method can not be applied to the the $\Phi$MAC blocks because the angles in every rotation gate in this block is a sum of angles in a range from the bigger to smaller angle, depending on the numbers to be multiplied (see Eq. \ref{eq:MultiplierAngles}).

\section{Conclusion}\label{sec:Conclusion}
\noindent
In this paper we have presented novel quantum arithmetic circuits, all based on the quantum Fourier transform representation of an integer or a superposition of integers; circuits are utilized for a novel, efficient realization of Shor's factorization algorithm.

The first circuit is a controlled multiplier by constant and accumulator ($\Phi$MAC) using $3n+1$ qubits and having a depth of $8n$, where $n$ is the bit width of the multiplication operands.

The second circuit (GM$\Phi$DIV) is a divider by a constant integer that produces both the quotient and the remainder. This circuit is inspired by an algorithm for classical computation given by Granlund and Montgomery \cite{Granlund}. The depth of the circuit in its second version (GM$\Phi$DIV2) is about $244n$ and it requires $7n+1$ qubits, where $n$ is the bit width of the quotient and the remainder.

The third circuit is a controlled modular multiplier, i.e. a circuit that multiplies a quantum integer by a constant and gives their product modulo another integer. Two versions of this circuit have been analyzed: 

\begin{itemize}
\item
a generic circuit ($\Phi$MUL\_MOD1) without restrictions in the range of integers it gets as input and constant. This circuit has a depth of about $2000n$, requires $9n+1$ qubits and has a total quantum cost of $2900n^{2}$ two-qubit gates; 
\item
an optimized circuit ($\Phi$MUL\_MOD2) suitable for Shor's algorithm with improved depth of about $1000n$, improved qubits requirement of $5n+1$ and a total quantum cost of $800n^{2}$ two-qubit gates. 
\end{itemize}

If we design the modular exponentiator block required by Shor's algorithm with this second optimized modular multiplier we can achieve a depth of $2000n^{2}$ using $9n+2$ qubits, where $n$ is the bit width of the integer to be factored by the algorithm. Further speed improvement can be achieved if we apply the approximate QFT method in every adder unit of the circuit. The proposed designs provide several advantages over other designs previously proposed in the literature in terms of circuit depth and qubits count.

\nonumsection{References}

\end{document}